\documentclass[10pt, journal, letterpaper]{IEEEtran}
\usepackage{url}
\usepackage[utf8]{inputenc}
\usepackage{amsmath}
\usepackage{amssymb}
\usepackage{xspace}
\usepackage{epsfig}
\usepackage{balance}
\usepackage{makecell}
\usepackage{listings}
\usepackage[dvipsnames]{xcolor}

\definecolor{codegray}{rgb}{0.25,0.25,0.25}
\definecolor{codepurple}{rgb}{0.58,0,0.82}
\lstdefinestyle{mystyle}{
  commentstyle=\color{PineGreen},
  keywordstyle=\color{MidnightBlue},
  numberstyle=\tiny\color{codegray},
  stringstyle=\color{codepurple},
  basicstyle=\ttfamily\footnotesize,
  breakatwhitespace=true,         
  breaklines=true,                 
  captionpos=b,
  frame=tb,
  keepspaces=true,                 
  numbers=left,                    
  numbersep=5pt,                  
  showspaces=false,                
  showstringspaces=false,
  showtabs=false,                  
  tabsize=2,
  xleftmargin=10pt,
  belowskip=-10pt,
  float=htbp,  
}

\lstdefinelanguage{mybash}{%
  language     = bash,
  morekeywords = {docker,python3},
}

\usepackage[acronyms,nonumberlist,nopostdot,nomain,nogroupskip,acronymlists={hidden}]{glossaries}
\newglossary[algh]{hidden}{acrh}{acnh}{Hidden Acronyms}

\usepackage{booktabs}
\usepackage{tabularx}

\usepackage{tikz}
\usepackage{pgfplots}
\pgfplotsset{compat=newest}
\pgfplotsset{plot coordinates/math parser=false}
\newlength\fheight
\newlength\fwidth
\usetikzlibrary{plotmarks,patterns,decorations.pathreplacing,backgrounds,calc,arrows,arrows.meta,spy,matrix,scopes}
\usepgfplotslibrary{patchplots,groupplots}
\usepackage{tikzscale}
\usepackage{hyperref}

\newif\ifexttikz
\exttikztrue

\ifexttikz
	\usetikzlibrary{external}
\fi

\usepackage{multirow}

\usepackage[font=footnotesize]{subcaption}
\usepackage[font=footnotesize]{caption}

\usepackage{mathtools}
\usepackage[numbers,sort&compress]{natbib}

\usepackage{soul}

\newacronym{3gpp}{3GPP}{3rd Generation Partnership Project}
\newacronym{4g}{4G}{4th generation}
\newacronym{5g}{5G}{5th generation}
\newacronym{6g}{6G}{6th generation}
\newacronym{5gc}{5GC}{5G Core}
\newacronym{adc}{ADC}{Analog to Digital Converter}
\newacronym{aerpaw}{AERPAW}{Aerial Experimentation and Research Platform for Advanced Wireless}
\newacronym{ai}{AI}{Artificial Intelligence}
\newacronym{aimd}{AIMD}{Additive Increase Multiplicative Decrease}
\newacronym{am}{AM}{Acknowledged Mode}
\newacronym{amc}{AMC}{Adaptive Modulation and Coding}
\newacronym{amf}{AMF}{Access and Mobility Management Function}
\newacronym{aops}{AOPS}{Adaptive Order Prediction Scheduling}
\newacronym{api}{API}{Application Programming Interface}
\newacronym{apn}{APN}{Access Point Name}
\newacronym{ap}{AP}{Application Protocol}
\newacronym{aqm}{AQM}{Active Queue Management}
\newacronym{ausf}{AUSF}{Authentication Server Function}
\newacronym{avc}{AVC}{Advanced Video Coding}
\newacronym{awgn}{AGWN}{Additive White Gaussian Noise}
\newacronym{balia}{BALIA}{Balanced Link Adaptation Algorithm}
\newacronym{bbu}{BBU}{Base Band Unit}
\newacronym{bdp}{BDP}{Bandwidth-Delay Product}
\newacronym{ber}{BER}{Bit Error Rate}
\newacronym{bf}{BF}{Beamforming}
\newacronym{bler}{BLER}{Block Error Rate}
\newacronym{brr}{BRR}{Bayesian Ridge Regressor}
\newacronym{bs}{BS}{Base Station}
\newacronym{bsr}{BSR}{Buffer Status Report}
\newacronym{bss}{BSS}{Business Support System}
\newacronym{ca}{CA}{Carrier Aggregation}
\newacronym{caas}{CaaS}{Connectivity-as-a-Service}
\newacronym{cb}{CB}{Code Block}
\newacronym{cc}{CC}{Congestion Control}
\newacronym{ccid}{CCID}{Congestion Control ID}
\newacronym{cco}{CC}{Carrier Component}
\newacronym{cdd}{CDD}{Cyclic Delay Diversity}
\newacronym{cdf}{CDF}{Cumulative Distribution Function}
\newacronym{cdn}{CDN}{Content Distribution Network}
\newacronym{cli}{CLI}{Command-line Interface}
\newacronym{cn}{CN}{Core Network}
\newacronym{codel}{CoDel}{Controlled Delay Management}
\newacronym{comac}{COMAC}{Converged Multi-Access and Core}
\newacronym{cord}{CORD}{Central Office Re-architected as a Datacenter}
\newacronym{cornet}{CORNET}{COgnitive Radio NETwork}
\newacronym{cosmos}{COSMOS}{Cloud Enhanced Open Software Defined Mobile Wireless Testbed for City-Scale Deployment}
\newacronym{cots}{COTS}{Commercial Off-the-Shelf}
\newacronym{cp}{CP}{Control Plane}
\newacronym{cyp}{CP}{Cyclic Prefix}
\newacronym{up}{UP}{User Plane}
\newacronym{cpu}{CPU}{Central Processing Unit}
\newacronym{cqi}{CQI}{Channel Quality Information}
\newacronym{cr}{CR}{Cognitive Radio}
\newacronym{cran}{CRAN}{Cloud \gls{ran}}
\newacronym{crs}{CRS}{Cell Reference Signal}
\newacronym{csi}{CSI}{Channel State Information}
\newacronym{csirs}{CSI-RS}{Channel State Information - Reference Signal}
\newacronym{cu}{CU}{Central Unit}
\newacronym{d2tcp}{D$^2$TCP}{Deadline-aware Data center TCP}
\newacronym{d3}{D$^3$}{Deadline-Driven Delivery}
\newacronym{dac}{DAC}{Digital to Analog Converter}
\newacronym{dag}{DAG}{Directed Acyclic Graph}
\newacronym{das}{DAS}{Distributed Antenna System}
\newacronym{dash}{DASH}{Dynamic Adaptive Streaming over HTTP}
\newacronym{dc}{DC}{Dual Connectivity}
\newacronym{dccp}{DCCP}{Datagram Congestion Control Protocol}
\newacronym{dce}{DCE}{Direct Code Execution}
\newacronym{dci}{DCI}{Downlink Control Information}
\newacronym{dctcp}{DCTCP}{Data Center TCP}
\newacronym{dl}{DL}{Downlink}
\newacronym{dmr}{DMR}{Deadline Miss Ratio}
\newacronym{dmrs}{DMRS}{DeModulation Reference Signal}
\newacronym{dnn}{DNN}{Deep Neural Network}
\newacronym{drlcc}{DRL-CC}{Deep Reinforcement Learning Congestion Control}
\newacronym{drs}{DRS}{Discovery Reference Signal}
\newacronym{du}{DU}{Distributed Unit}
\newacronym{e2e}{E2E}{end-to-end}
\newacronym{ecaas}{ECaaS}{Edge-Cloud-as-a-Service}
\newacronym{ecn}{ECN}{Explicit Congestion Notification}
\newacronym{edf}{EDF}{Earliest Deadline First}
\newacronym{embb}{eMBB}{Enhanced Mobile Broadband}
\newacronym{empower}{EMPOWER}{EMpowering transatlantic PlatfOrms for advanced WirEless Research}
\newacronym{enb}{eNB}{evolved Node Base}
\newacronym{endc}{EN-DC}{E-UTRAN-\gls{nr} \gls{dc}}
\newacronym{epc}{EPC}{Evolved Packet Core}
\newacronym{eps}{EPS}{Evolved Packet System}
\newacronym{es}{ES}{Edge Server}
\newacronym{etsi}{ETSI}{European Telecommunications Standards Institute}
\newacronym[firstplural=Estimated Times of Arrival (ETAs)]{eta}{ETA}{Estimated Time of Arrival}
\newacronym{eutran}{E-UTRAN}{Evolved Universal Terrestrial Access Network}
\newacronym{faas}{FaaS}{Function-as-a-Service}
\newacronym{fapi}{FAPI}{Functional Application Platform Interface}
\newacronym{fdd}{FDD}{Frequency Division Duplexing}
\newacronym{fdm}{FDM}{Frequency Division Multiplexing}
\newacronym{fdma}{FDMA}{Frequency Division Multiple Access}
\newacronym{fed4fire}{FED4FIRE+}{Federation 4 Future Internet Research and Experimentation Plus}
\newacronym{fir}{FIR}{Finite Impulse Response}
\newacronym{fit}{FIT}{Future \acrlong{iot}}
\newacronym{fpga}{FPGA}{Field Programmable Gate Array}
\newacronym{fr2}{FR2}{Frequency Range 2}
\newacronym{fs}{FS}{Fast Switching}
\newacronym{fscc}{FSCC}{Flow Sharing Congestion Control}
\newacronym{ftp}{FTP}{File Transfer Protocol}
\newacronym{fw}{FW}{Flow Window}
\newacronym{ge}{GE}{Gaussian Elimination}
\newacronym{gnb}{gNB}{Next Generation Node Base}
\newacronym{gop}{GOP}{Group of Pictures}
\newacronym{gpr}{GPR}{Gaussian Process Regressor}
\newacronym{gpu}{GPU}{Graphics Processing Unit}
\newacronym{gtp}{GTP}{GPRS Tunneling Protocol}
\newacronym{gtpc}{GTP-C}{GPRS Tunnelling Protocol Control Plane}
\newacronym{gtpu}{GTP-U}{GPRS Tunnelling Protocol User Plane}
\newacronym{gtpv2c}{GTPv2-C}{\gls{gtp} v2 - Control}
\newacronym{gw}{GW}{Gateway}
\newacronym{harq}{HARQ}{Hybrid Automatic Repeat reQuest}
\newacronym{hetnet}{HetNet}{Heterogeneous Network}
\newacronym{hh}{HH}{Hard Handover}
\newacronym{hol}{HOL}{Head-of-Line}
\newacronym{hqf}{HQF}{Highest-quality-first}
\newacronym{hss}{HSS}{Home Subscription Server}
\newacronym{http}{HTTP}{HyperText Transfer Protocol}
\newacronym{ia}{IA}{Initial Access}
\newacronym{iab}{IAB}{Integrated Access and Backhaul}
\newacronym{ic}{IC}{Incident Command}
\newacronym{ietf}{IETF}{Internet Engineering Task Force}
\newacronym{imsi}{IMSI}{International Mobile Subscriber Identity}
\newacronym{imt}{IMT}{International Mobile Telecommunication}
\newacronym{iot}{IoT}{Internet of Things}
\newacronym{ip}{IP}{Internet Protocol}
\newacronym{itu}{ITU}{International Telecommunication Union}
\newacronym{kpi}{KPI}{Key Performance Indicator}
\newacronym{kpm}{KPM}{Key Performance Measurement}
\newacronym{kvm}{KVM}{Kernel-based Virtual Machine}
\newacronym{los}{LOS}{Line-of-Sight}
\newacronym{lsm}{LSM}{Link-to-System Mapping}
\newacronym{lstm}{LSTM}{Long Short Term Memory}
\newacronym{lte}{LTE}{Long Term Evolution}
\newacronym{lxc}{LXC}{Linux Container}
\newacronym{m2m}{M2M}{Machine to Machine}
\newacronym{mac}{MAC}{Medium Access Control}
\newacronym{manet}{MANET}{Mobile Ad Hoc Network}
\newacronym{mano}{MANO}{Management and Orchestration}
\newacronym{mc}{MC}{Multi-Connectivity}
\newacronym{mcc}{MCC}{Mobile Cloud Computing}
\newacronym{mchem}{MCHEM}{Massive Channel Emulator}
\newacronym{mcs}{MCS}{Modulation and Coding Scheme}
\newacronym{mec}{MEC}{Multi-access Edge Computing}
\newacronym{mec2}{MEC}{Mobile Edge Cloud}
\newacronym{mfc}{MFC}{Mobile Fog Computing}
\newacronym{mgen}{MGEN}{Multi-Generator}
\newacronym{mi}{MI}{Mutual Information}
\newacronym{mib}{MIB}{Master Information Block}
\newacronym{miesm}{MIESM}{Mutual Information Based Effective SINR}
\newacronym{mimo}{MIMO}{Multiple Input, Multiple Output}
\newacronym{ml}{ML}{Machine Learning}
\newacronym{mlr}{MLR}{Maximum-local-rate}
\newacronym[plural=\gls{mme}s,firstplural=Mobility Management Entities (MMEs)]{mme}{MME}{Mobility Management Entity}
\newacronym{mmtc}{mMTC}{Massive Machine-Type Communications}
\newacronym{mmwave}{mmWave}{millimeter wave}
\newacronym{mpdccp}{MP-DCCP}{Multipath Datagram Congestion Control Protocol}
\newacronym{mptcp}{MPTCP}{Multipath TCP}
\newacronym{mr}{MR}{Maximum Rate}
\newacronym{mrdc}{MR-DC}{Multi \gls{rat} \gls{dc}}
\newacronym{mse}{MSE}{Mean Square Error}
\newacronym{mss}{MSS}{Maximum Segment Size}
\newacronym{mt}{MT}{Mobile Termination}
\newacronym{mtd}{MTD}{Machine-Type Device}
\newacronym{mtu}{MTU}{Maximum Transmission Unit}
\newacronym{mumimo}{MU-MIMO}{Multi-user \gls{mimo}}
\newacronym{mvno}{MVNO}{Mobile Virtual Network Operator}
\newacronym{nalu}{NALU}{Network Abstraction Layer Unit}
\newacronym{nas}{NAS}{Network Attached Storage}
\newacronym{nat}{NAT}{Network Address Translation}
\newacronym{nbiot}{NB-IoT}{Narrow Band IoT}
\newacronym{nfv}{NFV}{Network Function Virtualization}
\newacronym{nfvi}{NFVI}{Network Function Virtualization Infrastructure}
\newacronym{ni}{NI}{Network Interfaces}
\newacronym{nic}{NIC}{Network Interface Card}
\newacronym{nlos}{NLOS}{Non-Line-of-Sight}
\newacronym{now}{NOW}{Non Overlapping Window}
\newacronym{nsm}{NSM}{Network Service Mesh}
\newacronym[type=hidden]{nr}{NR}{New Radio}
\newacronym{nrf}{NRF}{Network Repository Function}
\newacronym{nsa}{NSA}{Non Stand Alone}
\newacronym{nse}{NSE}{Network Slicing Engine}
\newacronym{nssf}{NSSF}{Network Slice Selection Function}
\newacronym{o2i}{O2I}{Outdoor to Indoor}
\newacronym{oai}{OAI}{OpenAirInterface}
\newacronym{oaicn}{OAI-CN}{\gls{oai} \acrlong{cn}}
\newacronym{oairan}{OAI-RAN}{\acrlong{oai} \acrlong{ran}}
\newacronym{oam}{OAM}{Operations, Administration and Maintenance}
\newacronym{ofdm}{OFDM}{Orthogonal Frequency Division Multiplexing}
\newacronym{olia}{OLIA}{Opportunistic Linked Increase Algorithm}
\newacronym{omec}{OMEC}{Open Mobile Evolved Core}
\newacronym{onap}{ONAP}{Open Network Automation Platform}
\newacronym{onf}{ONF}{Open Networking Foundation}
\newacronym{onos}{ONOS}{Open Networking Operating System}
\newacronym{oom}{OOM}{\gls{onap} Operations Manager}
\newacronym{opnfv}{OPNFV}{Open Platform for \gls{nfv}}
\newacronym[type=hidden]{oran}{O-RAN}{Open \gls{ran}}
\newacronym{orbit}{ORBIT}{Open-Access Research Testbed for Next-Generation Wireless Networks}
\newacronym{os}{OS}{Operating System}
\newacronym{oss}{OSS}{Operations Support System}
\newacronym{pa}{PA}{Position-aware}
\newacronym{pase}{PASE}{Prioritization, Arbitration, and Self-adjusting Endpoints}
\newacronym{pawr}{PAWR}{Platforms for Advanced Wireless Research}
\newacronym{pbch}{PBCH}{Physical Broadcast Channel}
\newacronym{pcef}{PCEF}{Policy and Charging Enforcement Function}
\newacronym{pcfich}{PCFICH}{Physical Control Format Indicator Channel}
\newacronym{pcrf}{PCRF}{Policy and Charging Rules Function}
\newacronym{pdcch}{PDCCH}{Physical Downlink Control Channel}
\newacronym{pdcp}{PDCP}{Packet Data Convergence Protocol}
\newacronym{pdsch}{PDSCH}{Physical Downlink Shared Channel}
\newacronym{pdu}{PDU}{Packet Data Unit}
\newacronym{pf}{PF}{Proportional Fair}
\newacronym{pgw}{PGW}{Packet Gateway}
\newacronym{phich}{PHICH}{Physical Hybrid ARQ Indicator Channel}
\newacronym{phy}{PHY}{Physical}
\newacronym{pmch}{PMCH}{Physical Multicast Channel}
\newacronym{pmi}{PMI}{Precoding Matrix Indicators}
\newacronym{powder}{POWDER}{Platform for Open Wireless Data-driven Experimental Research}
\newacronym{ppo}{PPO}{Proximal Policy Optimization}
\newacronym{ppp}{PPP}{Poisson Point Process}
\newacronym{prach}{PRACH}{Physical Random Access Channel}
\newacronym{prb}{PRB}{Physical Resource Block}
\newacronym{psnr}{PSNR}{Peak Signal to Noise Ratio}
\newacronym{pss}{PSS}{Primary Synchronization Signal}
\newacronym{pucch}{PUCCH}{Physical Uplink Control Channel}
\newacronym{pusch}{PUSCH}{Physical Uplink Shared Channel}
\newacronym{qam}{QAM}{Quadrature Amplitude Modulation}
\newacronym{qci}{QCI}{\gls{qos} Class Identifier}
\newacronym{qoe}{QoE}{Quality of Experience}
\newacronym{qos}{QoS}{Quality of Service}
\newacronym{quic}{QUIC}{Quick UDP Internet Connections}
\newacronym{rach}{RACH}{Random Access Channel}
\newacronym{ran}{RAN}{Radio Access Network}
\newacronym[firstplural=Radio Access Technologies (RATs)]{rat}{RAT}{Radio Access Technology}
\newacronym{rbg}{RBG}{Resource Block Group}
\newacronym{rcn}{RCN}{Research Coordination Network}
\newacronym{rc}{RC}{RAN Control}
\newacronym{rec}{REC}{Radio Edge Cloud}
\newacronym{red}{RED}{Random Early Detection}
\newacronym{renew}{RENEW}{Reconfigurable Eco-system for Next-generation End-to-end Wireless}
\newacronym{rf}{RF}{Radio Frequency}
\newacronym{rfc}{RFC}{Request for Comments}
\newacronym{rfr}{RFR}{Random Forest Regressor}
\newacronym{ric}{RIC}{\gls{ran} Intelligent Controller}
\newacronym{rlc}{RLC}{Radio Link Control}
\newacronym{rlf}{RLF}{Radio Link Failure}
\newacronym{rlnc}{RLNC}{Random Linear Network Coding}
\newacronym{rmr}{RMR}{RIC Message Router}
\newacronym{rmse}{RMSE}{Root Mean Squared Error}
\newacronym{rnis}{RNIS}{Radio Network Information Service}
\newacronym{rr}{RR}{Round Robin}
\newacronym{rrc}{RRC}{Radio Resource Control}
\newacronym{rrm}{RRM}{Radio Resource Management}
\newacronym{rru}{RRU}{Remote Radio Unit}
\newacronym{rs}{RS}{Remote Server}
\newacronym{rsrp}{RSRP}{Reference Signal Received Power}
\newacronym{rsrq}{RSRQ}{Reference Signal Received Quality}
\newacronym{rss}{RSS}{Received Signal Strength}
\newacronym{rssi}{RSSI}{Received Signal Strength Indicator}
\newacronym{rtt}{RTT}{Round Trip Time}
\newacronym{ru}{RU}{Radio Unit}
\newacronym{rw}{RW}{Receive Window}
\newacronym{rx}{RX}{Receiver}
\newacronym{s1ap}{S1AP}{S1 Application Protocol}
\newacronym{sa}{SA}{standalone}
\newacronym{sack}{SACK}{Selective Acknowledgment}
\newacronym{sap}{SAP}{Service Access Point}
\newacronym{sc2}{SC2}{Spectrum Collaboration Challenge}
\newacronym{scef}{SCEF}{Service Capability Exposure Function}
\newacronym{sch}{SCH}{Secondary Cell Handover}
\newacronym{scoot}{SCOOT}{Split Cycle Offset Optimization Technique}
\newacronym{sctp}{SCTP}{Stream Control Transmission Protocol}
\newacronym{sdap}{SDAP}{Service Data Adaptation Protocol}
\newacronym{sdk}{SDK}{Software Development Kit}
\newacronym{sdm}{SDM}{Space Division Multiplexing}
\newacronym{sdma}{SDMA}{Spatial Division Multiple Access}
\newacronym{sdn}{SDN}{Software-defined Networking}
\newacronym{sdr}{SDR}{Software-defined Radio}
\newacronym{seba}{SEBA}{SDN-Enabled Broadband Access}
\newacronym{sgsn}{SGSN}{Serving GPRS Support Node}
\newacronym{sgw}{SGW}{Service Gateway}
\newacronym{si}{SI}{Study Item}
\newacronym{sib}{SIB}{Secondary Information Block}
\newacronym{sinr}{SINR}{Signal to Interference plus Noise Ratio}
\newacronym{sip}{SIP}{Session Initiation Protocol}
\newacronym{siso}{SISO}{Single Input, Single Output}
\newacronym{sla}{SLA}{Service Level Agreement}
\newacronym{sm}{SM}{Service Model}
\newacronym{smf}{SMF}{Session Management Function}
\newacronym{smo}{SMO}{Service Management and Orchestration}
\newacronym{sms}{SMS}{Short Message Service}
\newacronym{smsgmsc}{SMS-GMSC}{\gls{sms}-Gateway}
\newacronym{snr}{SNR}{Signal-to-Noise-Ratio}
\newacronym{son}{SON}{Self-Organizing Network}
\newacronym{sptcp}{SPTCP}{Single Path TCP}
\newacronym{srb}{SRB}{Service Radio Bearer}
\newacronym{srn}{SRN}{Standard Radio Node}
\newacronym{srs}{SRS}{Sounding Reference Signal}
\newacronym{ss}{SS}{Synchronization Signal}
\newacronym{sss}{SSS}{Secondary Synchronization Signal}
\newacronym{st}{ST}{Spanning Tree}
\newacronym{svc}{SVC}{Scalable Video Coding}
\newacronym{tb}{TB}{Transport Block}
\newacronym{tcp}{TCP}{Transmission Control Protocol}
\newacronym{tdd}{TDD}{Time Division Duplexing}
\newacronym{tdm}{TDM}{Time Division Multiplexing}
\newacronym{tdma}{TDMA}{Time Division Multiple Access}
\newacronym{tfl}{TfL}{Transport for London}
\newacronym{tfrc}{TFRC}{TCP-Friendly Rate Control}
\newacronym{tft}{TFT}{Traffic Flow Template}
\newacronym{tgen}{TGEN}{Traffic Generator}
\newacronym{tip}{TIP}{Telecom Infra Project}
\newacronym{tm}{TM}{Transparent Mode}
\newacronym{to}{TO}{Telco Operator}
\newacronym{tr}{TR}{Technical Report}
\newacronym{trp}{TRP}{Transmitter Receiver Pair}
\newacronym{ts}{TS}{Technical Specification}
\newacronym{tti}{TTI}{Transmission Time Interval}
\newacronym{ttt}{TTT}{Time-to-Trigger}
\newacronym{tx}{TX}{Transmitter}
\newacronym{uas}{UAS}{Unmanned Aerial System}
\newacronym{uav}{UAV}{Unmanned Aerial Vehicle}
\newacronym{udm}{UDM}{Unified Data Management}
\newacronym{udp}{UDP}{User Datagram Protocol}
\newacronym{udr}{UDR}{Unified Data Repository}
\newacronym{ue}{UE}{User Equipment}
\newacronym{uhd}{UHD}{\gls{usrp} Hardware Driver}
\newacronym{ul}{UL}{Uplink}
\newacronym{um}{UM}{Unacknowledged Mode}
\newacronym{uml}{UML}{Unified Modeling Language}
\newacronym{upa}{UPA}{Uniform Planar Array}
\newacronym{upf}{UPF}{User Plane Function}
\newacronym{urllc}{URLLC}{Ultra Reliable and Low Latency Communications}
\newacronym{usa}{U.S.}{United States}
\newacronym{usim}{USIM}{Universal Subscriber Identity Module}
\newacronym{usrp}{USRP}{Universal Software Radio Peripheral}
\newacronym{utc}{UTC}{Urban Traffic Control}
\newacronym{vim}{VIM}{Virtualization Infrastructure Manager}
\newacronym{vm}{VM}{Virtual Machine}
\newacronym{vnf}{VNF}{Virtual Network Function}
\newacronym{volte}{VoLTE}{Voice over \gls{lte}}
\newacronym{voltha}{VOLTHA}{Virtual OLT HArdware Abstraction}
\newacronym{vr}{VR}{Virtual Reality}
\newacronym{vran}{vRAN}{Virtualized \gls{ran}}
\newacronym{vss}{VSS}{Video Streaming Server}
\newacronym{wbf}{WBF}{Wired Bias Function}
\newacronym{wf}{WF}{Waterfilling}
\newacronym{wg}{WG}{Working Group}
\newacronym{wlan}{WLAN}{Wireless Local Area Network}
\newacronym{osm}{OSM}{Open Source \gls{nfv} Management and Orchestration}
\newacronym{pnf}{PNF}{Physical Network Function}
\newacronym{drl}{DRL}{Deep Reinforcement Learning}
\newacronym{mtc}{MTC}{Machine-type Communications}
\newacronym{osc}{OSC}{O-RAN Software Community}
\newacronym{mns}{MnS}{Management Services}
\newacronym{ves}{VES}{\gls{vnf} Event Stream}
\newacronym{ei}{EI}{Enrichment Information}
\newacronym{fh}{FH}{Fronthaul}
\newacronym{fft}{FFT}{Fast Fourier Transform}
\newacronym{laa}{LAA}{Licensed-Assisted Access}
\newacronym{plfs}{PLFS}{Physical Layer Frequency Signals}
\newacronym{ptp}{PTP}{Precision Time Protocol}
\tikzstyle{startstop} = [rectangle, rounded corners, minimum width=2cm, minimum height=0.5cm,text centered, draw=black]
\tikzstyle{io} = [trapezium, trapezium left angle=70, trapezium right angle=110, minimum width=3cm, minimum height=1cm, text centered, draw=black]
\tikzstyle{process} = [rectangle, minimum width=2cm, minimum height=0.5cm, text centered, draw=black, alignb=center]
\tikzstyle{decision} = [ellipse, minimum width=2cm, minimum height=1cm, text centered, draw=black]
\tikzstyle{arrow} = [thick,<->,>=stealth]
\tikzstyle{line} = [thick,>=stealth]
\tikzstyle{darrow} = [thick,<->,>=stealth,dashed]
\tikzstyle{sarrow} = [thick,->,>=stealth]
\tikzstyle{larrow} = [line width=0.1mm,dashdotted,->,>=stealth]
\tikzstyle{llarrow} = [line width=0.1mm,->,>=stealth]

\makeatletter
\def\grd@save@target#1{%
  \def\grd@target{#1}}
\def\grd@save@start#1{%
  \def\grd@start{#1}}
\tikzset{
  grid with coordinates/.style={
    to path={%
      \pgfextra{%
        \edef\grd@@target{(\tikztotarget)}%
        \tikz@scan@one@point\grd@save@target\grd@@target\relax
        \edef\grd@@start{(\tikztostart)}%
        \tikz@scan@one@point\grd@save@start\grd@@start\relax
        \draw[minor help lines] (\tikztostart) grid (\tikztotarget);
        \draw[major help lines] (\tikztostart) grid (\tikztotarget);
        \grd@start
        \pgfmathsetmacro{\grd@xa}{\the\pgf@x/1cm}
        \pgfmathsetmacro{\grd@ya}{\the\pgf@y/1cm}
        \grd@target
        \pgfmathsetmacro{\grd@xb}{\the\pgf@x/1cm}
        \pgfmathsetmacro{\grd@yb}{\the\pgf@y/1cm}
        \pgfmathsetmacro{\grd@xc}{\grd@xa + \pgfkeysvalueof{/tikz/grid with coordinates/major step x}}
        \pgfmathsetmacro{\grd@yc}{\grd@ya + \pgfkeysvalueof{/tikz/grid with coordinates/major step y}}
        \foreach \x in {\grd@xa,\grd@xc,...,\grd@xb}
        \node[anchor=north] at (\x,\grd@ya) {\pgfmathprintnumber{\x}};
        \foreach \y in {\grd@ya,\grd@yc,...,\grd@yb}
        \node[anchor=east] at (\grd@xa,\y) {\pgfmathprintnumber{\y}};
      }
    }
  },
  minor help lines/.style={
    help lines,
    gray,
    line cap =round,
    xstep=\pgfkeysvalueof{/tikz/grid with coordinates/minor step x},
    ystep=\pgfkeysvalueof{/tikz/grid with coordinates/minor step y}
  },
  major help lines/.style={
    help lines,
    line cap =round,
    line width=\pgfkeysvalueof{/tikz/grid with coordinates/major line width},
    xstep=\pgfkeysvalueof{/tikz/grid with coordinates/major step x},
    ystep=\pgfkeysvalueof{/tikz/grid with coordinates/major step y}
  },
  grid with coordinates/.cd,
  minor step x/.initial=.5,
  minor step y/.initial=.2,
  major step x/.initial=1,
  major step y/.initial=1,
  major line width/.initial=1pt,
}
\makeatother

\definecolor{desireRed}{RGB}{230,57,60}%
\definecolor{darkPurple}{RGB}{59,31,43}%
\definecolor{springGreen}{RGB}{37,223,145}%
\definecolor{queenBlue}{RGB}{69,123,157}%
\definecolor{spaceCadet}{RGB}{29,53,87}%

\usepackage{dblfloatfix}

\newcommand{\coloran}{ColO-RAN\xspace}
\newcommand{\openrangym}{OpenRAN Gym\xspace}
\newcommand{\scope}{SCOPE\xspace}

\usepackage[var0]{inconsolata}

\begin{document}
\bstctlcite{BSTcontrol}  

\title{\fontsize{23}{29}\selectfont OpenRAN Gym: AI/ML Development, Data~Collection, and Testing for O-RAN on PAWR Platforms}

\author{\IEEEauthorblockN{Leonardo Bonati, Michele Polese, Salvatore D'Oro, Stefano Basagni, Tommaso Melodia}\\
\IEEEauthorblockA{Institute for the Wireless Internet of Things, Northeastern University, Boston, MA, U.S.A.\\
E-mail: \{l.bonati, m.polese, s.doro, s.basagni, melodia\}@northeastern.edu}
\thanks{This is a revised and substantially extended version of~\cite{bonati2022openrangym}, which appeared in the Proceedings of the IEEE Wireless Communications and Networking Conference (WCNC) 2022 Workshops.}
\thanks{This work was partially supported by the U.S.\ National Science Foundation under Grants CNS-1925601, CNS-2120447, and CNS-2112471.}
}


\flushbottom
\setlength{\parskip}{0ex plus0.1ex}

\maketitle
\glsunset{nr}
\glsunset{lte}
\glsunset{3gpp}

\begin{abstract}
Open \gls{ran} architectures will enable interoperability, openness and programmable data-driven control in next generation cellular networks.
However, developing and testing efficient solutions that generalize across heterogeneous cellular deployments and scales, and that optimize network performance in such diverse environments is a complex task that is still largely unexplored.
In this paper we present \openrangym, a unified, open, and O-RAN-compliant experimental toolbox for data collection, design, prototyping and testing of end-to-end data-driven control solutions for next generation Open \gls{ran} systems.
\openrangym extends and combines into a unique solution several software frameworks for data collection of \gls{ran} statistics and \gls{ran} control, and a lightweight O-RAN near-real-time \gls{ric} tailored to run on experimental wireless platforms.
We first provide an overview of the various architectural components of \openrangym and describe how it is used to collect data and design, train and test artificial intelligence and machine learning O-RAN-compliant applications (xApps) at scale.
We then describe in detail how to test the developed xApps on softwarized \glspl{ran} and provide an example of two xApps developed with \openrangym that are used to control a network with~7 base stations and~42 users deployed on the Colosseum testbed.
Finally, we show how solutions developed with \openrangym on Colosseum can be exported to real-world, heterogeneous wireless platforms, such as the Arena testbed and the POWDER and COSMOS platforms of the PAWR program.
\openrangym and its software components are open-source and publicly-available to the research community.
By guiding the readers from instantiating the components of \openrangym, to running experiments in a softwarized \gls{ran} with an O-RAN-compliant near-RT \gls{ric} and xApps, we aim at providing a key reference for researchers and practitioners working on experimental Open \gls{ran} systems.
\end{abstract}

\begin{picture}(0,0)(10,-530)
\put(0,0){
\put(0,0){\footnotesize This work has been published on the Elsevier Computer Networks journal \url{https://doi.org/10.1016/j.comnet.2022.109502}.}
\put(0,-10){
\footnotesize\!\!\textcopyright 2022. This manuscript version is made available under the CC-BY-NC-ND 4.0 license.}}
\end{picture}

\glsresetall
\glsunset{nr}
\glsunset{lte}
\glsunset{3gpp}

\section{Introduction}

%
Once seen as monolithic and mostly immutable ``black-box'' systems, cellular networks are converging toward the more flexible, software-based open architectures based on the Open \gls{ran} paradigm.
This new approach to cellular communications promotes openness, virtualization, and programmability of  \gls{ran} functionalities and components,
and enables data-driven intelligent control loops for cellular systems~\cite{bonati2020open}.
As such, the Open \gls{ran} enables network operators to support new bespoke services on shared physical infrastructures, and to dynamically reconfigure them based on network conditions and user demand.
The resulting increased efficiency will also decrease the operational costs of the network.

In this context, standardization bodies and other organizations are releasing a number of specifications to regulate the operations of the Open \gls{ran}, and to define its capabilities, constraints, and use cases.
The most notable is the O-RAN Alliance, which is developing specifications---collected under the O-RAN umbrella---to apply Open \gls{ran} principles to prevailing radio access technologies, including \gls{3gpp} LTE and NR networks~\cite{oran-wg1-arch-spec}.

O-RAN introduces two network \glspl{ric}, operating at different timescales, enabling programmatic closed-loop control of the \gls{ran} elements. 
It also defines a set of open interfaces to connect the controllers to key elements of the \gls{ran}, such as the NR \glspl{cu}, \glspl{du}, \glspl{ru}, and the LTE O-RAN-compliant \glspl{enb}~\cite{polese2022understanding}.
In details, the near-real-time (or near-RT) \gls{ric} connects to the \gls{ran} elements (i.e., the \glspl{cu} and \glspl{du}) through the E2 interface, and enables control loops operating at timescales ranging between~$10$\:ms and~$1$\:s~\cite{oran-wg3-ricarch}.
Instead, the non-real-time (or non-RT) \gls{ric} is included as part of \gls{smo} frameworks, and operates at timescales larger than $1$\:s~\cite{oran-wg2-non-rt-ric-architecture}.
This component also interacts with one or multiple near-RT \glspl{ric} via the A1 interface, which is used to disseminate policies and information external to the network.
The non-RT \gls{ric} also manages the \gls{ai} and \gls{ml} models which are instantiated on the \glspl{ric} in the form of standalone applications, namely xApps (on the near-RT \gls{ric}) and rApps (on the non-RT \gls{ric}).
Finally, the \gls{smo} connects to the \gls{ran} through the O1 interface, used for management and orchestration routines, and to the O-RAN virtualization platform (the O-Cloud) via the O2 interface.

\begin{figure*}[t]
\setlength\belowcaptionskip{-10pt}
    \centering
    \includegraphics[width=.9\textwidth]{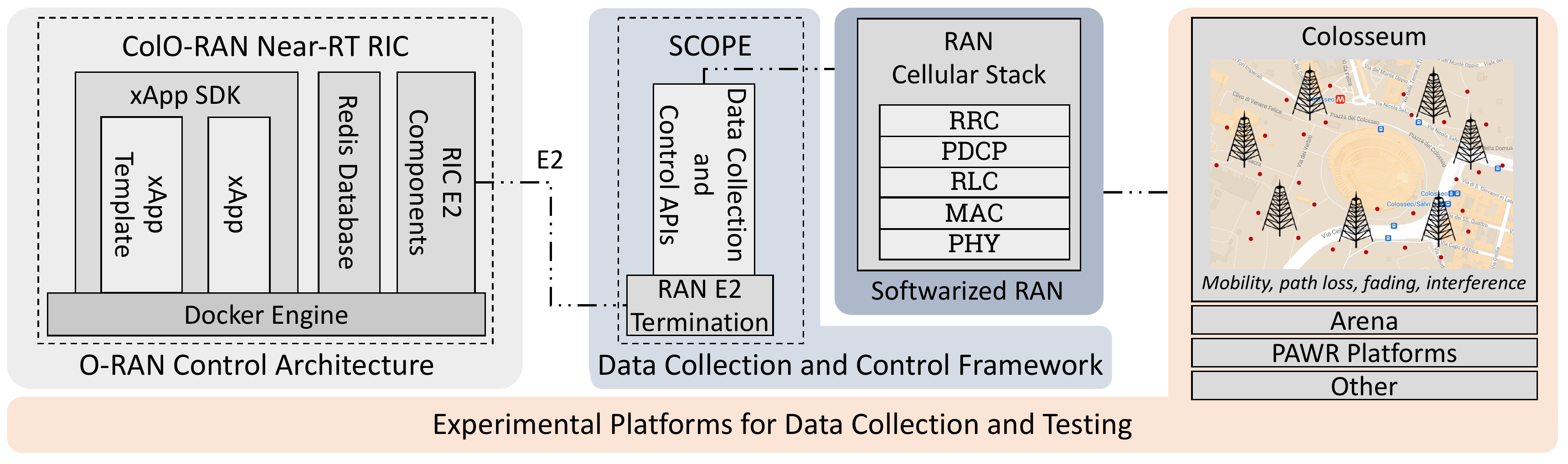}
    \caption{\openrangym architecture.}
    \label{fig:openran-architecture}
\end{figure*}

Thanks to its \glspl{ric}, open interfaces and disaggregated architecture, O-RAN ultimately enables the practical deployment and execution of \gls{ai}/\gls{ml} solutions at scale, which can be used to infer and forecast network traffic, or to reconfigure the nodes of the \gls{ran} at run time based on real-time conditions and user demand.
Typical workflows for the design and testing of such \gls{ai}/\gls{ml} algorithms encompass a number of different steps such as~\cite{oran-wg2-ml,polese2021coloran}):
(i)~\emph{data collection}, to create practical datasets representative of the different environments (e.g., the wireless channel) where the \gls{ai}/\gls{ml} models will be deployed, as well as of various performance indicators of the network;
(ii)~\emph{AI/ML model design}, selecting the inputs and outputs of the models, and \emph{training and testing}, to evaluate the effectiveness and limits of such models;
(iii)~\emph{model deployment} as applications deployed on the \glspl{ric}, i.e., xApps/rApps or---as recently proposed in \cite{doro2022dapps}---directly on the \glspl{cu}/\glspl{du} via dApps;
(iv)~\emph{model fine-tuning} with run-time data from the \gls{ran}, to adapt the models to different production environments,
and (v)~the actual \emph{control, inference and/or forecasting} or the \gls{ran}.

In this paper we present \openrangym, an open-source toolbox to develop \gls{ai}/\gls{ml} O-RAN-compliant inference and control algorithms, to deploy them as xApps on the near-RT \gls{ric}, and to test them on a large-scale softwarized \gls{ran} controlled by the \gls{ric}.
Throughout this work, we guide readers on how to start using the \openrangym framework to run experiments in a softwarized Open \gls{ran} managed by an O-RAN-compliant near-RT \gls{ric}, how to implement custom \gls{ai}/\gls{ml} solutions of their design as xApps on the Colosseum wireless network emulator, and how to transition such solution on over-the-air open testbeds for wireless research.
First, we give a high-level overview of the various components of \openrangym, and discuss how they enable development and testing workflows of data-driven xApps.
We showcase an example of two xApps designed with \openrangym and used to control a large-scale \gls{ran} instantiated on the Colosseum wireless network emulator~\cite{bonati2021colosseum} through the \scope framework~\cite{bonati2021scope}, and controlled by the \coloran near-RT \gls{ric}~\cite{polese2021coloran}.
%
We also show how \openrangym can be seamlessly ported from an emulator such as Colosseum to over-the-air real-world platforms, such as the Arena testbed~\cite{bertizzolo2020arena}, and the platforms of the U.S.\ National Science Foundation-sponsored \gls{pawr} program~\cite{pawr} including the \gls{powder}~\cite{breen2021powder} and the \gls{cosmos}~\cite{raychaudhuri2020cosmos} platforms.
To the best of our knowledge, \openrangym is the first open, portable toolset for end-to-end design, prototyping, testing, and experimentation of \gls{ai}/\gls{ml} O-RAN xApps on heterogeneous wireless experimental platforms.
As such, we hope that this work can be an important reference for researchers and practitioners working on---or starting to work on---experimental Open \gls{ran} systems.

Previous experimental work has focused on the development of data-driven solutions and xApps for specific use cases~\cite{s21248173,johnson2021open}, on the description of the \gls{ai}/\gls{ml} capabilities of O-RAN~\cite{lee2020hosting,abdalla2021generation}, on interoperability testing~\cite{oran2019plugfest}, and on orchestration~\cite{doro2022orchestran}. Compared to the state of the art, \openrangym enables an end-to-end workflow for the design and testing of \gls{ai}/\gls{ml} solutions as xApps in the O-RAN ecosystem.
By doing so, it empowers users with a first-of-its-kind open and publicly-available O-RAN-compliant toolbox that will unleash the potential of data-driven applications for next generation cellular networks.
\openrangym aims at contributing to the thriving community of wireless researchers and developers by providing open-source software components for experimental O-RAN-enabled data-driven research.
%
We actively maintain an up-to-date resource on the \openrangym project\footnote{\url{https://openrangym.com}} that can be used to review the functionalities of our framework, and as a reference to repositories, documentation, tutorials, and containers. This also includes publications that describe in details each component, the different use cases in which they were used, as well as public datasets collected through \openrangym.

The remainder of this paper is organized as follows. 
We give an overview of the various components of \openrangym in Section~\ref{sec:openrangym}.
Practical descriptions of \openrangym data collection and control framework, and O-RAN control architecture, are given in Sections~\ref{sec:scope} and~\ref{sec:coloran}, respectively.
The xApp design and testing workflow is presented in Section~\ref{sec:xapp-design-testing-workflow}, along with an example of large-scale \gls{ran} control using xApps developed with \openrangym on Colosseum.
Section~\ref{sec:traveling} discusses how \openrangym components and experiments can be ported from Colosseum to heterogeneous real-world testbeds.
Section~\ref{sec:results} showcases exemplary results obtained on the different platforms considered in this work.
Finally, conclusions are drawn in Section~\ref{sec:conclusions}.

\section{OpenRAN Gym}
\label{sec:openrangym}

The \openrangym architecture is shown in Figure~\ref{fig:openran-architecture}.
Its main components are: (i)~publicly- and remotely-accessible \textit{experimental wireless platforms} for collecting data, prototyping, and testing solutions in heterogeneous environments. Example of these are the Colosseum wireless network emulator~\cite{bonati2021colosseum}, the Arena testbed~\cite{bertizzolo2020arena}, and the platforms of the PAWR program~\cite{pawr}; (ii)~a \textit{softwarized \gls{ran}} implemented through open protocol stacks for cellular networks, such as srsRAN~\cite{gomez2016srslte} and OpenAirInterface~\cite{kaltenberger2020openairinterface}; (iii)~a \textit{data collection and control framework}, such as \scope~\cite{bonati2021scope}, that exposes \glspl{api} to extract relevant \glspl{kpm} from the \gls{ran}, and dynamically control it at run-time, and (iv)~an \textit{O-RAN control architecture}, such as \coloran~\cite{polese2021coloran}, able to connect to the \gls{ran} through open and standardized interfaces (e.g., the O-RAN E2 interface), receive the run-time \glspl{kpm} from the \gls{ran}, and control it through \gls{ai}/\gls{ml} solutions running, for instance, as xApps/rApps.
As we will show in Sections~\ref{sec:traveling} and~\ref{sec:results}, \openrangym is platform-independent, and it allows users to perform data collection campaigns, prototype, and evaluate solutions in a set of heterogeneous wireless environments and deployments before transitioning them to production networks.
As such, \openrangym can be used to first prototype and validate solutions on the Colosseum wireless network emulator, and then seamlessly transfer such solutions to heterogeneous platforms, such as the Arena testbed, and the \acrshort{powder} and \acrshort{cosmos} platforms from the \gls{pawr} program~\cite{pawr}.
The procedures to port the various components of \openrangym on these platforms will be described in Section~\ref{sec:traveling}.

Arena is an indoor wireless testbed equipped with a grid of 64~antennas and 24~\glspl{sdr} (among USRPs~X310 and~N210) controlled by high-performance compute servers~\cite{bertizzolo2020arena}.
Its deployment is representative of a live office environment.

Colosseum is the world's largest wireless network emulator~\cite{bonati2021colosseum}.
It allows researchers and practitioners to experiment at scale, and in different channel conditions and virtual environments through a set of 128~\glspl{sdr} (USRPs~X310) controlled through dedicated servers---namely, \glspl{srn}---interconnected through a \gls{mchem}.
The latter, is capable of reproducing conditions of the wireless channel (e.g., path loss, fading, user mobility, signal interference and superimposition) by means of \gls{fir} filters implemented through \glspl{fpga}.
The channel emulation is performed by the \gls{fir} filters, which apply the channel impulse response of the desired wireless channel to the  signals transmitted by the \glspl{srn}.
Sets of channel impulse responses for different environments (e.g., urban, rural, etc.)---referred to as \textit{\gls{rf} scenarios} in Colosseum---are modeled a priori through mathematical equations, or captured through ray-tracing software.

\gls{powder} is a city-scale wireless testbed deployed in Salt Lake City, UT~\cite{breen2021powder}.
The testbed includes a number of \glspl{sdr} deployed across an outdoor area, an over-the-air indoor laboratory setup, and a wired attenuator matrix.
The objective of this testbed is to foster experimental research in heterogeneous technology, such as 5G cellular technologies and network orchestration.

\gls{cosmos} is a city-scale testbed deployed in New York City, NY, which mainly focuses on mmWave communications with edge-computing capabilities~\cite{raychaudhuri2020cosmos}.
This testbed absorbed the \gls{orbit}~\cite{kohli2021openaccess}, an indoor over-the-air wireless platform with remotely-accessible \gls{sdr} devices and compute servers.

At the time of this writing, \openrangym softwarized \gls{ran} leverages the cellular implementation provided by srsRAN~\cite{gomez2016srslte}, which allows users to instantiate protocol stacks of \gls{3gpp} base stations and \glspl{ue} using \glspl{sdr} as front-end interfaces.
This cellular protocol stack is augmented by the \scope framework, which adds a number of networking and control functionalities to srsRAN including network slicing capabilities, support for additional scheduling algorithms, data collection pipelines, and open \glspl{api} to control such functionalities at run time.
As we will discuss in Section~\ref{sec:scope}, \scope can facilitate data collection campaigns by automating the collection of relevant \gls{ran} \gls{kpm} in the heterogeneous testbed where it is instantiated~\cite{bonati2021intelligence,polese2021coloran,doro2022orchestran}.

Finally, \coloran implements the O-RAN control architecture of \openrangym.
This framework adapts the near-RT \gls{ric} provided by the \gls{osc} to run in a lightweight containerized environment, and extends it to swiftly interface with, and control, the \scope base stations through the E2 interface standardized by O-RAN.
As we will discuss in Section~\ref{sec:coloran}, \coloran allows users to prototype \gls{ai}/\gls{ml}-based O-RAN applications through an \textit{xApp \gls{sdk}}, to instantiate them on an \gls{osc}-compliant near-RT \gls{ric}, and to leverage them to perform control of a softwarized \gls{ran} (Figure~\ref{fig:openran-architecture}).

\section{Data Collection and Control Framework}
\label{sec:scope}

The data collection and control framework of \openrangym is based on \scope~\cite{bonati2021scope}.
This framework provides a programmable environment for prototyping and testing solutions for softwarized \glspl{ran}, and data collection capabilities of relevant \glspl{kpm} (e.g., throughput, \glspl{tb}, buffer occupancy).
Concerning the cellular protocol stack for base stations and \glspl{ue}, \scope leverages srsRAN~\cite{gomez2016srslte}, which it extends with novel network slicing and a set of additional scheduling algorithms.
Open \glspl{api} to fine-tune the configuration of the \gls{ran} at run time, and to perform data collection campaigns are also provided by \scope.
Coupled with different testbeds---such as Colosseum and the platform of the PAWR program---\scope can facilitate the collection of \gls{ran} \glspl{kpm} in a set of heterogeneous scenarios and environments by automatically collect such statistics from the running experiments~\cite{bonati2021scope,bonati2021intelligence,polese2021coloran}.
%
Finally, \scope connects to the O-RAN near-RT \gls{ric} through a \gls{ran}-side O-RAN E2 termination, which is based on the \gls{osc} \gls{du}~\cite{osc-du-l2}.
This allows user-defined xApps running on the near-RT \gls{ric} to swiftly interface with the \gls{ran} base stations, and to dynamically control their functionalities at run time (e.g., modify the scheduling policy and set the amount of resources allocated to each network slice).
In the remainder of this section, we will give a high-level overview of the main configuration options and parameters of the \scope-enabled base stations, and show how to instantiate a cellular network with it.
\scope has been open-sourced to the research community,\footnote{The \scope source code is available at \url{https://github.com/wineslab/colosseum-scope} and \url{https://github.com/wineslab/colosseum-scope-e2}.} and also provided to the Colosseum users in the form of a ready-to-use \gls{lxc} (namely \texttt{scope}/\texttt{scope-with-e2}).
In Section~\ref{sec:traveling}, we will show how the publicly-available \scope container can be ported to different testbeds (e.g., the Arena testbed, and the \gls{powder} and \gls{cosmos} testbeds of the PAWR program) with minor modifications.
In this way, \scope truly enables the process of cellular-network-as-a-service, in which the solutions are first prototyped in a controlled environment (e.g., Colosseum), and then ported in the wild on real-world testbeds.

\subsection{Starting \scope}
\label{sec:scope-setup}

\scope provides \gls{cli} tools to start the cellular base stations and configure them through parameters passed via configuration files.
The main parameters of interest to \openrangym are described as follows.\footnote{A comprehensive description of the \scope \glspl{api} and configuration parameters can be found at \url{https://github.com/wineslab/colosseum-scope}.}

\begin{itemize}
\item \texttt{network-slicing}: enables/disables the network slicing functionalities of the base station.

\item \texttt{slice-allocation}: if network slicing has been enabled, this parameter can be used to set the \glspl{rbg} allocated by the base station to each slice. The input of this configuration option is passed as \texttt{\{slice:[first\_rbg, last\_rbg],...\}}.
As an example, \texttt{\{0:[0,5],1:[6,10]\}} allocates \glspl{rbg}~0-5 to slice~0 and~6-10 to slice~1.

\item \texttt{slice-scheduling-policy}: sets the scheduling policy used for each network slice of the base station.
As an example, \texttt{[1,2]} assigns slicing policy~1 to slice~0 and policy~2 to slice~1.
The possible numerical values for this field match the scheduling policies supported by \scope (i.e., \texttt{0}: round-robin, \texttt{1}: waterfilling, \texttt{2}: proportionally fair).

\item \texttt{slice-users}: associates \glspl{ue} to a specific network slice. The input of this configuration option is passed as \texttt{\{slice:[ue1,ue2],...\}}.
As an example, \texttt{\{0:[4,5],1:[2,3]\}} assigns \glspl{ue}~4, 5 to slice~0, and \glspl{ue}~2, 3 to slice~1.
%
%

\item \texttt{generic-testbed}: specifies whether \scope is running on a testbed other than Colosseum.
In this case, the parameters \texttt{node-is-bs} and \texttt{ue-id} can also be passed to specify whether the node should act as a base station or a \gls{ue}, and the identifier of the \gls{ue} in the latter case.\footnote{When running on Colosseum, \scope automatically derives the role of the node, and the \gls{ue} identifier based on the allocated \glspl{srn}.}

\end{itemize}

After the \scope configuration has been written in a JSON-formatted file,\footnote{An example of configuration file can be found at \url{https://tinyurl.com/2s3pvw83} (for Colosseum), and at \url{https://tinyurl.com/35t7s97a} (for testbeds other than Colosseum).}
(named \texttt{radio.conf} in the code snippet below), the cellular base station, core network, and \gls{ue} applications can be started
through the commands of Listing~\ref{lst:scope-start}.
\begin{lstlisting}[language=mybash,style=mystyle,
caption={Commands to start the \scope applications.},
label={lst:scope-start}]
#!/bin/bash
cd radio_api/
python3 scope_start.py --config-file radio.conf
\end{lstlisting}

At run-time, the \scope \glspl{api} can be leveraged to fine-tune the configuration of the base station, e.g., to modify the scheduling policy of each slice, or to set the amount of \glspl{rbg} of each slice (see~\cite[Section~3.3]{bonati2021scope}).
Relevant \glspl{kpm} from the \gls{ran} are automatically logged by the \scope base stations while traffic is exchanged among base stations and \glspl{ue}.
These \glspl{kpm} are saved in CSV-formatted files that 
%
can be either used on-the-fly (e.g., for online \gls{ai}/\gls{ml} model training or inference), or retrieved for offline processing after the experiment ends (e.g., to perform offline \gls{ai}/\gls{ml} model training).

\section{O-RAN Control Architecture}
\label{sec:coloran}

The O-RAN control architecture leveraged by \openrangym is based on \coloran, an open-source framework to develop, design, prototype, and test O-RAN-ready solutions at scale~\cite{polese2021coloran}.
This framework provides a lightweight implementation of the \gls{osc} near-RT \gls{ric}---which has been adapted to run on the Colosseum system as a set of standalone Docker containers---as well as automated pipelines for the deployment of the various services of the \gls{ric}.
The main components implemented by \coloran are shown in Figure~\ref{fig:openran-architecture}, left.
They are services in charge of overseeing the interactions with the \gls{ran} (e.g., the \textit{E2 termination, E2 manager,} and \textit{E2 routing manager}), a \textit{Redis database} that keeps records of the connected \gls{ran} nodes (e.g., the base stations), and an \textit{xApp \gls{sdk}} with tools to prototype and test \gls{ai}/\gls{ml}-based xApp for run-time \gls{ran} inference and/or control.

\begin{figure}[ht]
    \centering
    \includegraphics[width=\columnwidth]{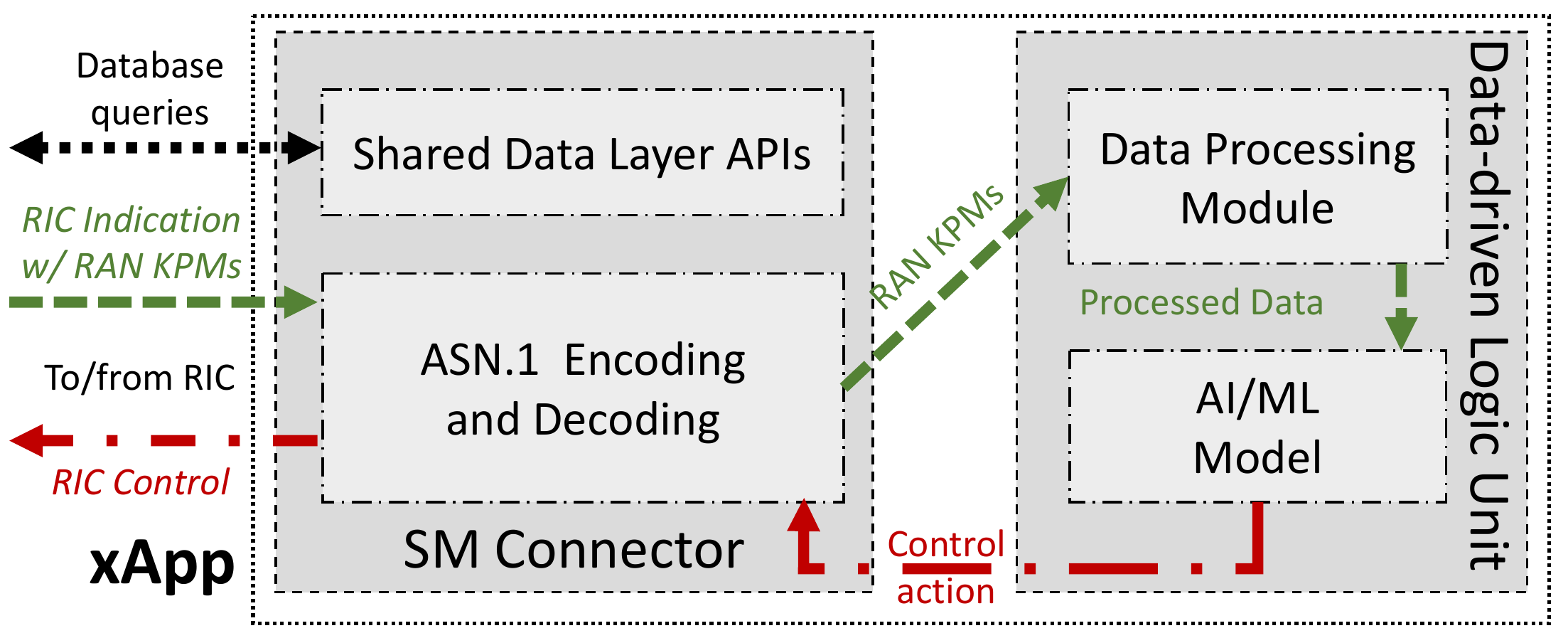}
    \caption{\coloran xApp, adapted from~\cite{polese2021coloran}.}
    \label{fig:xapp}
\end{figure}

A high-level diagram of a \coloran xApp is shown in Figure~\ref{fig:xapp}.
This is formed of two main parts: (i)~the \textit{\gls{sm} connector}, in charge of handling the messages between the xApp and the near-RT \gls{ric} (e.g., the control messages for the \gls{ran}), and (ii)~the \textit{data-driven logic unit}, which processes \glspl{kpm} received from the \gls{ran} base stations, and performs tasks based on \gls{ai}/\gls{ml} models (e.g., traffic prediction and/or control of the functionalities of the base stations).
The data-driven logic unit hosts two sub-components, namely the \textit{\gls{ai}/\gls{ml} model} and the \textit{data processing module}.
The former consists of the specific data-driven model (e.g., \gls{drl} agent, \gls{dnn}, \gls{lstm}, to name a few) used to perform inference and/or control tasks.
The latter, instead, executes data processing functionalities to convert the input \glspl{kpm} into data that can be fed to the \gls{ai}/\gls{ml} model.
For instance, the majority of \gls{ai}/\gls{ml} models are designed to receive inputs with a fixed size and format (e.g., a two-dimensional array of a specific length, an image, or a time-series). However, the \glspl{kpm} received over the E2 termination might have a different format, or might contain more data than what is required by the \gls{ai}/\gls{ml} model. In this case, the data processing module performs the necessary operations to convert the input data in the correct format.
In some cases, the data processing module can also host some \gls{ai}/\gls{ml} models that execute advanced data processing operations.
Examples of these are autoencoders to extract latent data representation and to perform dimensionality reduction~\cite{polese2021coloran,bonati2021intelligence}. 

In the remainder of this section, we detail how to instantiate the \coloran near-RT \gls{ric}
(Section~\ref{sec:near-rt-ric-start}), how to interface it with the \scope base station through the O-RAN E2 termination (Section~\ref{sec:scope-ric-connect}), and how to start a sample xApp that controls the base station (Section~\ref{sec:setting-xapp}).
\coloran has been open-sourced and made available to the research community,\footnote{The \coloran source code is available at \url{https://github.com/wineslab/colosseum-near-rt-ric}.} and also provided to the Colosseum users in the form of a ready-to-use \gls{lxc} container (namely \texttt{coloran-near-rt-ric}).
In Section~\ref{sec:traveling}, we will show how \coloran can be seamlessly ported to different testbeds (e.g., the Arena testbed, and the \gls{powder} and \gls{cosmos} platforms of the \gls{pawr} program).

\subsection{Starting the \coloran Near-RT RIC}
\label{sec:near-rt-ric-start}

The \coloran near-RT \gls{ric} can be built and instantiated as a set of Docker containers by running the \texttt{setup-ric.sh} script and the commands of Listing~\ref{lst:ric-start}.\footnote{A pre-built version of \coloran, named \texttt{coloran-near-rt-ric-prebuilt}, is also provided to the Colosseum users. The pre-built Docker images are also hosted on Docker Hub at \url{https://hub.docker.com/u/wineslab}.}
This script, adapted from~\cite{johnson-powder-ric-profile}, takes as input the network interface the \gls{ric} uses to receive and exchange messages with the \gls{ran} (e.g., the \texttt{col0} interface in Colosseum).
\begin{lstlisting}[language=mybash,style=mystyle,
caption={Commands to set up the \coloran near-RT \gls{ric}.},
label={lst:ric-start}]
#!/bin/bash
cd setup-scripts/
./setup-ric.sh col0
\end{lstlisting}
As a first step, the base Docker images that will be used to build the \gls{ric} are imported.
Then, the actual Docker images of the \coloran near-RT \gls{ric} are built.
These images include: (i)~the \texttt{e2term}, which is the endpoint of the \gls{ric} E2 messages;
%
(ii)~the \texttt{e2mgr}, which is in charge of managing the messages to/from the E2 interface;
%
(iii)~the \texttt{e2rtmansim}, which uses the \gls{rmr} protocol to route the E2 messages within the \gls{ric};
%
and (iv)~the \texttt{db}, which implements a Redis database with records of the \gls{ran} nodes connected to the \gls{ric} (e.g., the base stations).
%
During this step, the IP addresses and ports that will be used by the Docker containers are also configured as set up in the \texttt{setup-lib.sh} file.
After the Docker images have been built, the \gls{ric} containers, listening for incoming connections from the \gls{ran} through the E2 termination endpoint, are spawned.
The logs of the various containers can be accessed through the \texttt{docker logs} command, e.g., \texttt{docker logs -f e2term} shows the run-time logs of the E2 termination (\texttt{e2term}) container.

\subsection{Connecting the \scope Base Station to \coloran}
\label{sec:scope-ric-connect}

After setting up and starting \coloran through the steps described in Section~\ref{sec:near-rt-ric-start}, the cellular base station---provided by \scope and set up in Section~\ref{sec:scope-setup}---can be connected to it through the O-RAN E2 termination, which has been adapted from the \gls{osc} \gls{du} implementation~\cite{osc-du-l2}.
To this aim, the \gls{ran}-side E2 termination can be used to: (i)~receive \textit{\gls{ric} Subscription} messages from the xApps; (ii)~transmit periodic \gls{kpm} reports to the xApps through \textit{\gls{ric} Indication} messages, and receive control actions from them through \textit{\gls{ric} Control} messages, and (iii)~interact with the \glspl{api} provided by \scope to modify the configuration of the base station at run time (e.g., the scheduling and slicing policies) based on the control messages received from the xApps.
The steps to initialize the E2 termination at the \scope base station are shown in Listing~\ref{lst:odu-start}.
\begin{lstlisting}[language=mybash,style=mystyle,
%aboveskip=10pt,
caption={Commands initialize the \scope E2 termination process.},
label={lst:odu-start}]
#!/bin/bash
cd colosseum-scope-e2/
./build_odu.sh clean
./run_odu.sh
\end{lstlisting}
First, the E2 termination is built through the \texttt{build\_odu.sh} script (line~3).
This script also specifies the IP address and port of the near-RT \gls{ric} to connect to, as well as the local network interface used for the connection.
Then, the E2 termination process is started through the \texttt{run\_odu.sh} script (line~4), which establishes the initial connection between the base station and the near-RT \gls{ric}.
The successful outcome of this connection can be verified in the logs of the \texttt{e2term} container (via the \texttt{docker logs -f e2term} command, see Section~\ref{sec:near-rt-ric-start}), which reports the identifier of the connected base station (e.g., \texttt{gnb:311-048-01000501}).

\begin{figure*}[t]
\setlength\belowcaptionskip{-10pt}
    \centering
    \includegraphics[width=.94\textwidth]{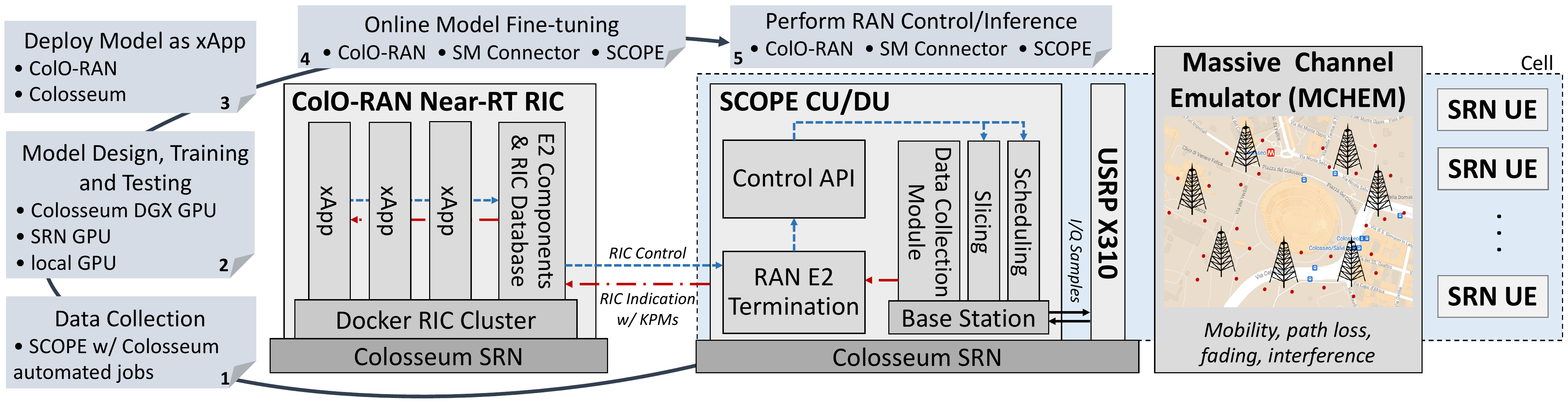}
    \caption{\openrangym xApp design and testing workflow on Colosseum.}
    \label{fig:xapp-workflow}
\end{figure*}

\subsection{Initializing a Sample xApp}
\label{sec:setting-xapp}

After the \scope base station has been connected to the near-RT \gls{ric}, the xApps can be started.
To facilitate the design of novel xApps, we provide a ready-to-use sample xApp template in which researchers and practitioners can plug-in their custom \gls{ai}/\gls{ml} models. 
This sample xApp can be started through the \texttt{setup-sample-xapp.sh} script and the commands shown in Listing~\ref{lst:xapp-build}.
\begin{lstlisting}[language=mybash,style=mystyle,
caption={Commands to build the \coloran sample xApp Docker image, and to start and configure the xApp container.},
label={lst:xapp-build}]
#!/bin/bash
cd setup-scripts/
./setup-sample-xapp.sh gnb:311-048-01000501
\end{lstlisting}
This script takes as input the identifier of the base station the xApp should subscribe to (see Section~\ref{sec:scope-ric-connect}), and builds the Docker image of the sample xApp.
Then, the script starts the xApp Docker container---dubbed \texttt{sample-xapp}---on the near-RT \gls{ric}.

After the container has started, the xApp processes can be run through the commands of Listing~\ref{lst:xapp-run}.
\begin{lstlisting}[language=mybash,style=mystyle,
caption={Commands to run the \coloran sample xApp process.},
label={lst:xapp-run}]
#!/bin/bash
docker exec -it sample-xapp /home/sample-xapp/run_xapp.sh
\end{lstlisting}
These commands trigger the xApp subscription to the targeted \gls{ran} nodes (e.g., one or multiple base stations connected to the \gls{ric}) through \gls{ric} Subscription messages, and the periodic reports of \gls{ran} \glspl{kpm}
from such nodes.
Starting from the provided template, \openrangym users can build xApps running custom solutions (e.g., with custom \gls{ai}/\gls{ml} agents).

\section{xApp Development Workflow on Colosseum}
\label{sec:xapp-design-testing-workflow}

The main steps to develop a data-driven xApp using \openrangym on Colosseum are shown in Figure~\ref{fig:xapp-workflow}.

1)~\textbf{Data collection.}
This step involves collecting the data that will be used to train and test the \gls{ai}/\gls{ml} model to embed in the xApp.
In Colosseum, this can be done by combining the data-collection capabilities of \scope with the automated experiments of Colosseum.
This allows to automatically run experiments with several base stations and users in a set of heterogeneous scenarios, and to collect the \gls{ran} \glspl{kpm}---saved in CSV-formatted files---from Colosseum \gls{nas} once the experiment ends~\cite{bonati2021scope}.
%
%
%
%

2)~\textbf{Model design, training and testing.}
After data collection campaigns have been performed in heterogeneous wireless environments and scenarios, the \gls{ai}/\gls{ml} model can be designed.
This step includes the selection of the \gls{ai}/\gls{ml} algorithm that the model will use, along with the data used as input, the reward function, and the set of output actions (e.g., to perform inference or control of the \gls{ran}).
After this design phase, the model is first trained offline using the data collected in step~1, and then tested at scale.\footnote{It is worth mentioning that the O-RAN specifications forbid the deployment of models that have not been trained offline beforehand. This is to shield the \gls{ran} from poor performance or outages~\cite{oran-wg2-ml}.}
Being computationally-intensive, this step may benefit from \gls{gpu}-enabled environments.
As such, they can be carried out either locally (i.e., on the user's own \gls{gpu}-enabled machines), or on Colosseum's \gls{gpu}-enabled \glspl{srn} or NVIDIA A100 DGXs.

3)~\textbf{Deploy the model as an xApp.}
After the model has been tested (step~2), it can be deployed as an xApp on the \coloran near-RT \gls{ric} by following the procedures of Section~\ref{sec:setting-xapp}.
Specifically, the \gls{ai}/\gls{ml} model is included in the \textit{data-driven logic unit} of \coloran xApp (see Figure~\ref{fig:xapp}) by modifying the provided xApp template.
Finally, the modified xApp is built and instantiated on the near-RT \gls{ric} through the commands of Listings~\ref{lst:xapp-build} and~\ref{lst:xapp-run}.

4)~\textbf{Online model fine-tuning.}
At run-time, the xApp communicates with the \scope base station through the near-RT \gls{ric} and the E2 termination.
To this aim, the xApp first subscribes to the base station by sending it a \gls{ric} Subscription message.
Then, it triggers periodic \glspl{kpm} reports---with periodicity tunable based on the needs of the users~\cite{polese2021coloran}---from the base station.
These reports are sent through \gls{ric} Indication messages, and they may be used by the xApp to fine-tune the model online, allowing it to adapt to varying wireless conditions and traffic demand.
Once the model has been fine-tuned online, the Docker image of the xApp can be updated with the trained weights.

5)~\textbf{Perform \gls{ran} control/inference.}
At this stage, the xApp can be used in the a live infrastructure to perform inference and/or control of the \gls{ran}.
This entails the xApp transmitting the actions computed by the model to the \scope base station through \gls{ric} Control messages.
Example of these are actions to modify the parameters and configuration of the base station, e.g., to modify the resources allocated to the slices of the network, or their scheduling policies.
At the base station, these \gls{ric} Control messages---received through the O-RAN E2 interface---trigger the \scope control \glspl{api} of Figure~\ref{fig:xapp-workflow}, which apply the new policies to the configuration of the base station at run time.
At this point, the xApps can be tested and validated on Colosseum.
In Sections~\ref{sec:traveling} and~\ref{sec:results}, we will show how these newly developed xApps can be ported and instantiated on external wireless testbeds.

\subsection{Example of xApps} \label{sec:xapps:examples}

For the sake of completeness, we now provide an example of two xApps designed, trained and tested with \openrangym on Colosseum.
These xApps are used to control a cellular network with 7~base stations and 42~\glspl{ue}
instantiated on the Colosseum network emulator.
Base stations adopt a \gls{fdd} configuration with $50$~\glspl{prb} (corresponding to $10$\:MHz of bandwidth).
Each base station is implemented through \scope and serves 6~\glspl{ue} with different traffic requirements.
The \glspl{ue} are divided into two classes of traffic, allocated to different slices of the network: time-sensitive (e.g, \gls{urllc}) and broadband (e.g., \gls{embb} and \gls{mtc}).
Further examples of xApps developed with \openrangym and used to optimize the network performance of a softwarized \gls{ran} instantiated on a general-purpose infrastructure are discussed in details in~\cite{bonati2021intelligence,polese2021coloran,doro2022orchestran,bonati2022intelligent}.

\begin{figure}[t]
    \centering
    \includegraphics[width=\columnwidth]{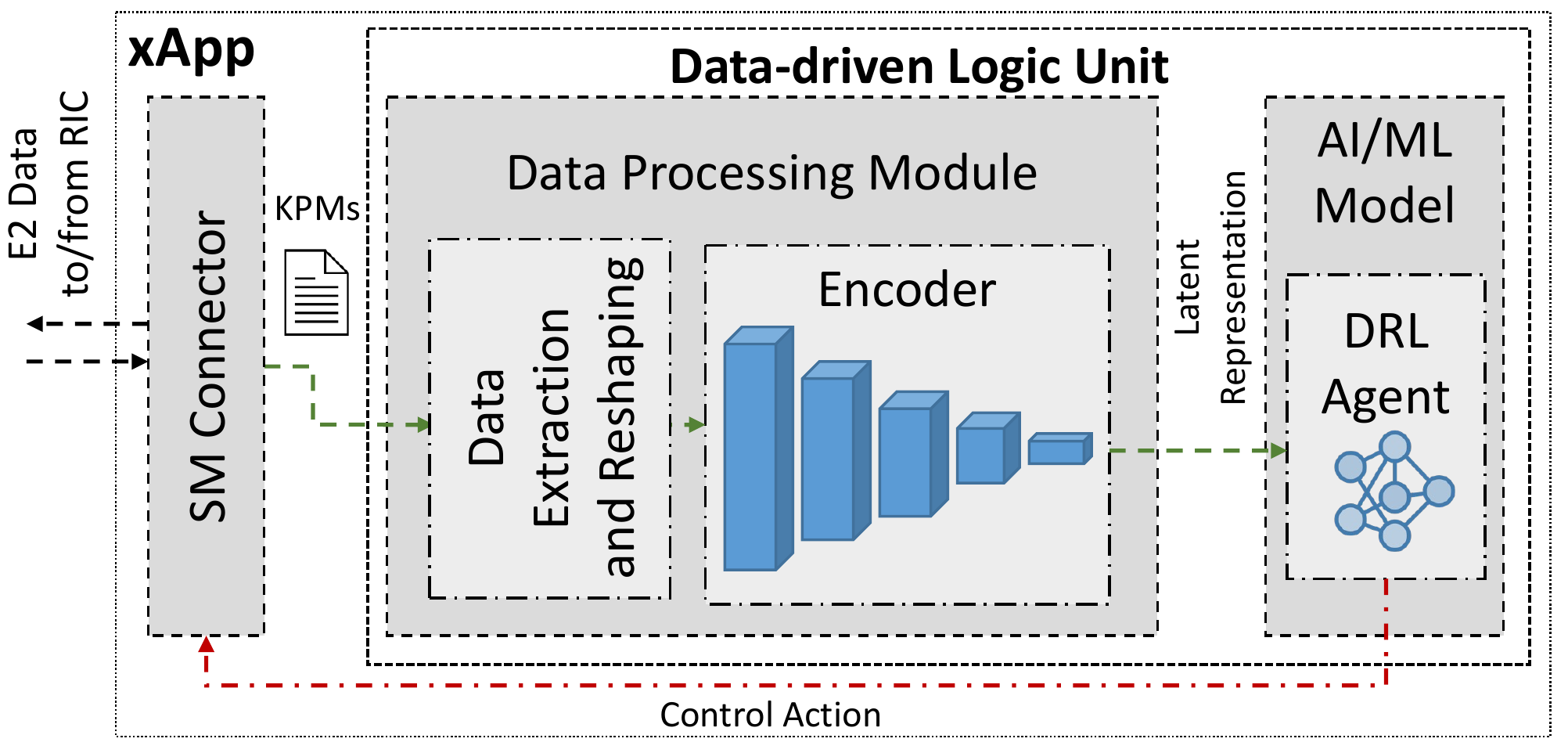}
    \caption{The architecture of the xApps used in Section~\ref{sec:xapps:examples}. Adapted from~\cite{bonati2022intelligent}.}
    \label{fig:xapps_example}
\end{figure}

%
The xApp structure used in this section is shown in Figure~\ref{fig:xapps_example}. Although the general architecture stems from that depicted in Figure~\ref{fig:xapp}, in this specific example, the data-driven logic unit consists of two elements: (i) an encoder for data dimensionality reduction, which is in charge of converting \gls{kpm} reports received on the E2 interface into a latent (and low dimension) representation of the data, and (ii) a \gls{drl} agent that converts the latent representation into a state of the network and computes the optimal control action that maximizes an agent-specific reward according to the  current state.

\textbf{Agent Design.}~DRL agents are trained on a dataset with $3.4$\:GB of \gls{ran} traces (and more than 73~hours of experiments) collected on Colosseum---that make control decisions on the configuration of the base station based on the received \gls{ran} \glspl{kpm} (see~\cite{polese2021coloran}).
The \gls{drl} agents considered in this paper implement a \gls{ppo} architecture that leverages an actor-critic structure.
The actor network is trained to take actions according to the current state of the system, while the critic network is used during the training phase to evaluate the reward obtained by selecting a specific action in a certain state.
Then, the critic network instructs the actor network on how valuable the action was, in this way steering the actor network toward actions that bring the highest reward for each state. Both actor and critic networks consist of fully-connected neural networks with 5 layers, where each layer has 30 neurons.

\textbf{Actions.}~To showcase the impact of different design choices on the overall performance of the network, we trained two xApps with different action spaces.
One xApp (named \texttt{sched}) controls the scheduling policies that a base station uses for specific classes of traffic. 
Another xApp (\texttt{sched-slicing}) is instead operating over a larger action space as it controls both scheduling policies, and the resource allocated to each slice (i.e., the number of \glspl{rbg} assigned to each class of traffic).

\textbf{Reward.}~In this example, both xApps aim at maximizing the amount of transmitted data belonging to the broadband traffic class (in this case measured by the number of downlink \glspl{tb} transmitted successfully by the base station to the \glspl{ue}), and minimizing the end-to-end latency of the time-sensitive traffic class.
As the protocol stack of the base station does not have a direct measurement of the end-to-end system latency, we use the buffer occupancy metric as our proxy for latency, which reflects how much time each packet spends in the transmission buffer queue.

To capture these objectives, the reward of the \gls{drl} agents consists of a reward function that jointly maximizes the throughput for the broadband slice and minimizes the downlink buffer size for the time-sensitive slice. These two elements are combined with a weighted sum, whose details can be found in~\cite{polese2021coloran,bonati2022intelligent}.

\begin{figure}[ht]
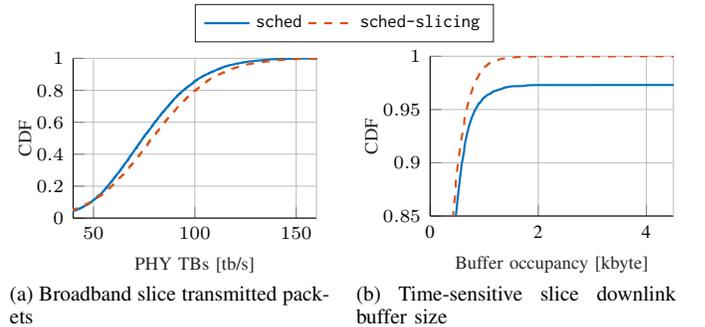

\setlength\abovecaptionskip{0.1cm}
\setlength\belowcaptionskip{-1pt}
    \centering
    \ifexttikz
        \tikzsetnextfilename{sched-vs-sched-slicing-pkt-slice-1}
    \fi
    \begin{subfigure}[b]{0.48\columnwidth}
        \centering
        \setlength\fwidth{.8\columnwidth}
        \setlength\fheight{.5\columnwidth}
        \input{figures/sched-vs-sched-slicing-pkt-slice-1.tex}
       \setlength\abovecaptionskip{-0.4cm}
        \caption{Broadband slice transmitted packets}
        \label{fig:xapp-pkts-mtc}
    \end{subfigure}\hfill
    \ifexttikz
        \tikzsetnextfilename{sched-vs-sched-slicing-buffer-slice-2}
    \fi
    \begin{subfigure}[b]{0.48\columnwidth}
        \centering
        \setlength\fwidth{.8\columnwidth}
        \setlength\fheight{.5\columnwidth}
        \input{figures/sched-vs-sched-slicing-buffer-slice-2.tex}
       \setlength\abovecaptionskip{-0.4cm}
        \caption{Time-sensitive slice downlink buffer size}
        \label{fig:xapp-buffer-urllc}
    \end{subfigure}\hfill
    \caption{Comparison of xApps developed with \openrangym.}
    \label{fig:xapp-comparison}
\end{figure}

Figure~\ref{fig:xapp-comparison} shows the \gls{cdf} of some \gls{ran} metrics measured at the base stations when the two xApps are instantiated on the \coloran near-RT \gls{ric} and used to control the \gls{ran}.
Specifically, Figure~\ref{fig:xapp-pkts-mtc} shows the transmitted \glspl{tb} for the broadband slice, while Figure~\ref{fig:xapp-buffer-urllc} displays the downlink buffer occupancy of the time-sensitive slice.
By acting on a large action set (i.e., the slice resource allocation), the \texttt{sched-slicing} xApp achieves superior performance by delivering a higher number of transmitted packets and reducing the occupancy of the downlink buffer.\footnote{A detailed evaluation of \openrangym xApps, including their orchestration, and control of large-scale experimental networks can be found in~\cite{polese2021coloran, doro2022orchestran}.}

\begin{table*}[ht]
\setlength\belowcaptionskip{5pt}
    \centering
    \footnotesize
    \setlength{\tabcolsep}{2pt}
    \caption{Compute node and radio setups used across the different testbeds.}
    \label{tab:compute-node-setup}
    \begin{tabularx}{\textwidth}{
        >{\raggedright\arraybackslash\hsize=\hsize}X
        >{\raggedright\arraybackslash\hsize=1.5\hsize}X
        >{\raggedright\arraybackslash\hsize=\hsize}X
        >{\raggedleft\arraybackslash\hsize=0.65\hsize}X
        >{\raggedleft\arraybackslash\hsize=0.65\hsize}X
        >{\raggedleft\arraybackslash\hsize=1.2\hsize}X }
        \toprule
        Testbed & Compute Node & Processor & CPU Cores & RAM [GB] & Software-defined Radio \\
        \midrule
        \multicolumn{6}{>{\centering\arraybackslash\hsize=\dimexpr6\hsize+6\tabcolsep+\arrayrulewidth\relax}X}{Base Station (BS) / UE} \\
        \midrule
        Arena       & Dell PowerEdge R340           & Intel Xeon E-2146G    & 6   & 32    & NI USRP~X310 \\
        Colosseum   & Dell PowerEdge R730           & Intel Xeon E5-2650    & 48  & 128   & NI USRP~X310 \\
        COSMOS      & Asus server (model undisclosed)   & Intel i7-4790         & 4   & 16    & NI USRP~B210 \\
        POWDER (BS) & Dell PowerEdge R740           & Intel Xeon Gold 6126  & 24  & 98    & NI USRP~X310 \\
        POWDER (UE) & Intel NUC 8559                    & Intel 7-8559U         & 4   & 32    & NI USRP~B210 \\
        \midrule
        \multicolumn{6}{>{\centering\arraybackslash\hsize=\dimexpr6\hsize+6\tabcolsep+\arrayrulewidth\relax}X}{Near-RT RIC} \\
        \midrule
        Arena       & Dell PowerEdge R340   & Intel Xeon E-2146G    & 6   & 32    & N/A \\
        Colosseum   & Dell PowerEdge R730   & Intel Xeon E5-2650    & 48  & 128   & N/A \\
        COSMOS      & Supermicro 1028U-TRT+     & Intel Xeon E5-2698    & 16  & 251   & N/A \\
        POWDER      & Dell PowerEdge R740   & Intel Xeon Gold 6126  & 24  & 98    & N/A \\
        \bottomrule
    \end{tabularx}
\end{table*}

\section{Traveling Containers}
\label{sec:traveling}


In this section, we illustrate how the \openrangym containerized applications including the xApps developed and pre-trained on Colosseum can be transferred to other testbeds, and describe the necessary adjustments (if any) to run these applications on each experimental platform.
Although the above procedure may seem trivial in the case of self-contained applications, in our case this is challenging due to the fact that our \textit{traveling \openrangym containers} need to interact with the underlying network resources and be able to properly control the potentially diverse set of \glspl{sdr} available in the different testbeds.
Furthermore, in some cases, the firmware of the \glspl{sdr} requires specific tools only available on certain operating systems versions or distributions. In such cases, the containers may need to be updated or rebuilt.

To facilitate these tasks, we developed some tools to automatically start the \openrangym \gls{lxc} containers on the different platforms considered in this work, and to properly interface them with the available radio resources.
After the \gls{lxc} images have been transferred (e.g., through the \texttt{scp} or \texttt{rsync} utilities) in a running instance of the testbed of interest,
the image can be imported with the commands shown in Listing~\ref{lst:lxc-image-import}.
\begin{lstlisting}[language=mybash,style=mystyle,
caption={Commands to import the \scope LXC image with the E2 termination module.},
label={lst:lxc-image-import}]
#!/bin/bash
lxc image import scope-with-e2.tar.gz --alias scope-e2
\end{lstlisting}
This command imports the \texttt{scope-with-e2.tar.gz} \gls{lxc} image transferred from Colosseum (i.e., the \scope image with the module for the E2 termination) to the compute machine of the remote testbed.
%
After the above operation completes successfully, the new image, named \texttt{scope}, is visible by running the following command: \texttt{lxc image list}.

The \gls{lxc} container can be, then, created from the imported image by running the commands shown in Listing~\ref{lst:lxc-container-create}.
\begin{lstlisting}[language=mybash,style=mystyle,
caption={Commands to create the \scope LXC container with the E2 termination module from the image imported in Listing~\ref{lst:lxc-image-import}.},
label={lst:lxc-container-create}]
#!/bin/bash
lxc init local:scope-e2 scope
\end{lstlisting}

After creating the container, additional operations may be required based on the specific \openrangym image, and \gls{sdr} available in the remote testbed (e.g., USRP~B210 or~X310).
As an example, if running the \scope container with an USRP~B210, it is necessary to perform an \textit{USB passthrough} operation to allow the container to use the USB interfaces and devices connected to the physical host (e.g., the USB interface to control the USRP).
If using an USRP~X310, instead, the container needs access to the network interface the host machine uses to communicate with the \gls{sdr}, to set the right \gls{mtu} for it, and possibly to flash the \gls{fpga} of the USRP with the appropriate image.
In both these cases, the container may require some additional permissions to be able to use the passed devices and interfaces (e.g., read/write permissions on the USB devices).

In the case of \coloran---which can be imported and started with commands analogous to those shown in Listings~\ref{lst:lxc-image-import} and~\ref{lst:lxc-container-create}---it is necessary to configure the \gls{nat} rules of the host machine for it to forward the messages directed to the \gls{ric} to the \coloran container (e.g., the \textit{E2 Setup Request} message used by the base station to subscribe to the \gls{ric}, and the \textit{\gls{ric} Indication} messages used to send the \glspl{kpm} to the xApps).
Similarly to the previous case, the \gls{lxc} container may require some additional permissions (e.g., to run the Docker containers of the \coloran near-RT \gls{ric} in a \textit{nested} manner inside the \coloran \gls{lxc} container).

After these operations have been executed, the \gls{lxc} container (e.g., the \scope \gls{lxc} container created in Listing~\ref{lst:lxc-container-create}) can be started with the commands of Listing~\ref{lst:lxc-container-start}.
\begin{lstlisting}[language=mybash,style=mystyle,
caption={Commands to start the \scope LXC container created in Listing~\ref{lst:lxc-container-create}.},
label={lst:lxc-container-start}]
#!/bin/bash
lxc start scope
\end{lstlisting}
Now, the \openrangym applications can be executed by following the procedures detailed in Sections~\ref{sec:scope-setup}, \ref{sec:near-rt-ric-start}, \ref{sec:scope-ric-connect}, and~\ref{sec:setting-xapp}.

\smallskip
To simplify and automate the above setup operations, and to allow \openrangym users to swiftly configure and run the transferred containers, we developed and open-sourced a set of scripts that take care of (i)~passing the right radio interface to the containers; (ii)~giving them the required permissions; (iii)~setting up the \gls{nat} rules of the host machine, and, finally, (iv) starting the \openrangym \gls{lxc} containers from the imported images~\footnote{\url{https://github.com/wineslab/openrangym-pawr}}.
These scripts, which are supposed to be run after the commands of Listing~\ref{lst:lxc-image-import}, i.e., after the \gls{lxc} image has been imported, are described in Listings~\ref{lst:scope-container-start} and~\ref{lst:coloran-container-start}.

Specifically, Listing~\ref{lst:scope-container-start} creates, sets up, and starts the \scope \gls{lxc} container starting from the image imported in Listing~\ref{lst:lxc-image-import}.
\begin{lstlisting}[language=mybash,style=mystyle,
caption={Commands to start the \scope LXC container.},
label={lst:scope-container-start}]
#!/bin/bash
./start-lxc-scope.sh testbed usrp_type [flash]
\end{lstlisting}
After creating the container on the \texttt{testbed} of interest (i.e., \texttt{arena}, \texttt{powder}, or \texttt{cosmos}), the script configures the USRP specified through the \texttt{usrp\_type} parameter (i.e., \texttt{b210} or \texttt{x310}) following the procedures described above (i.e., passing to the container the devices to interface with the USRP, and assigning the appropriate permissions to the container).
%
The optional \texttt{flash} parameter also allows to flash the \gls{fpga} of the USRP~X310 with the UHD image used by the container.\footnote{Please note that after the \gls{fpga} has been flashed with a new image, the USRP may need to be rebooted. We refer to the documentation of the various testbeds for the instructions on how to achieve this.}
%

The script of Listing~\ref{lst:coloran-container-start} can be used to create, setup, and start the \coloran near-RT \gls{ric} container starting from the image imported in Listing~\ref{lst:lxc-image-import}.
\begin{lstlisting}[language=mybash,style=mystyle,
caption={Commands to start the \coloran LXC container.},
label={lst:coloran-container-start}]
#!/bin/bash
./start-lxc-ric.sh
\end{lstlisting}
This script creates the \coloran near-RT \gls{ric} container, assigns it the required permissions (e.g., to run the nested Docker containers), and starts it.
Then, it sets the \gls{nat} rules of the host machine (where the container is running) for it to forward the messages intended for the \gls{ric}.
Finally, it builds and starts the Docker containers of the \coloran near-RT \gls{ric} (see Section~\ref{sec:coloran}) inside the created \gls{lxc} container.

\begin{figure*}[t]
\centering
\begin{minipage}[t]{0.48\textwidth}
  \centering
  \ifexttikz
      \tikzsetnextfilename{colosseum-stairs}
  \fi
  \begin{subfigure}[t]{0.48\columnwidth}
    \setlength\fwidth{.8\columnwidth}
    \setlength\fheight{.55\columnwidth}
    \input{figures/colosseum-stairs.tex}
    \caption{Colosseum}
    \label{fig:colosseum-stairs}
  \end{subfigure}\hfill
  \ifexttikz
      \tikzsetnextfilename{arena-stairs}
  \fi
  \begin{subfigure}[t]{0.48\columnwidth}
    \setlength\fwidth{.8\columnwidth}
    \setlength\fheight{.55\columnwidth}
    \input{figures/arena-stairs.tex}
    \caption{Arena}
    \label{fig:arena-stairs}
  \end{subfigure}
  \ifexttikz
      \tikzsetnextfilename{powder-stairs}
  \fi
  \begin{subfigure}[t]{0.48\columnwidth}
    \setlength\fwidth{.8\columnwidth}
    \setlength\fheight{.55\columnwidth}
    \input{figures/powder-stairs.tex}
    \caption{POWDER}
    \label{fig:powder-stairs}
  \end{subfigure}\hfill
  \ifexttikz
      \tikzsetnextfilename{cosmos-stairs}
  \fi
  \begin{subfigure}[t]{0.48\columnwidth}
    \setlength\fwidth{.8\columnwidth}
    \setlength\fheight{.55\columnwidth}
    \input{figures/cosmos-stairs.tex}
    \caption{COSMOS}
    \label{fig:cosmos-stairs}
  \end{subfigure}
  \caption{Overall slice throughput varying the percentage of \glspl{rbg} allocated to each slice over time according to the configuration reported in Table~\ref{tab:slice-configuration}.}
  \label{fig:slice-stairs}
\end{minipage}
\hfill
\begin{minipage}[t]{0.48\textwidth}
  \centering
  \ifexttikz
      \tikzsetnextfilename{colosseum-v}
  \fi
  \begin{subfigure}[t]{0.48\columnwidth}
    \setlength\fwidth{.8\columnwidth}
    \setlength\fheight{.55\columnwidth}
    \input{figures/colosseum-v.tex}
    \caption{Colosseum}
    \label{fig:colosseum-v}
  \end{subfigure}\hfill
  \ifexttikz
      \tikzsetnextfilename{arena-v}
  \fi
  \begin{subfigure}[t]{0.48\columnwidth}
    \setlength\fwidth{.8\columnwidth}
    \setlength\fheight{.55\columnwidth}
    \input{figures/arena-v.tex}
    \caption{Arena}
    \label{fig:arena-v}
  \end{subfigure}\hfill
  \ifexttikz
      \tikzsetnextfilename{powder-v}
  \fi
  \begin{subfigure}[t]{0.48\columnwidth}
    \setlength\fwidth{.8\columnwidth}
    \setlength\fheight{.55\columnwidth}
    \input{figures/powder-v.tex}
    \caption{POWDER}
    \label{fig:powder-v}
  \end{subfigure}
  \ifexttikz
      \tikzsetnextfilename{cosmos-v}
  \fi
  \begin{subfigure}[t]{0.48\columnwidth}
    \setlength\fwidth{.8\columnwidth}
    \setlength\fheight{.55\columnwidth}
    \input{figures/cosmos-v.tex}
    \caption{COSMOS}
    \label{fig:cosmos-v}
  \end{subfigure}
  \caption{Overall slice throughput varying the percentage of \glspl{rbg} allocated to each slice over time according to the configuration reported in Table~\ref{tab:slice-configuration}.}
  \label{fig:slice-v}
\end{minipage}
\end{figure*}

\section{Experimental Results}
\label{sec:results}

In this section, we showcase some experimental results obtained from running \openrangym and its components across a set of heterogeneous testbeds.
We ported the \scope and \coloran near-RT \gls{ric} containers from Colosseum to the Arena, \gls{powder}, and \gls{cosmos} testbeds (see Sections~\ref{sec:openrangym} and~\ref{sec:traveling}).
A description of the
setups used in these testbeds (also summarized in Table~\ref{tab:compute-node-setup}) follows.
Since the capabilities offered by the different testbeds can be substantially different (e.g., number of available over-the-air nodes), for the sake of consistency, and to fairly compare results, we run experiments with one cellular base station and up to three \glspl{ue}, and one near-RT \gls{ric} node.
In all cases, we use a \gls{fdd} configuration with $50$~\glspl{prb}, corresponding to $10$\:MHz of bandwidth.
We divide the spectrum of the base stations into up to three network slices, and statically assign the \glspl{ue} to them (e.g., based on the \gls{sla} between users and their network operator).
Downlink \gls{udp} traffic generated through the iPerf3 tool is leveraged to evaluate the network performance.
Finally, the base stations---implemented through \scope---connect to \coloran near-RT \gls{ric} through the E2 interface standardized by O-RAN.

\textbf{POWDER.}
We instantiated both the \coloran near-RT \gls{ric} and the \scope base station on Dell PowerEdge R740 compute nodes,
%
while the \glspl{ue} were instantiated on Intel NUC 8559 nodes.
%
The radio front-end of the base station was implemented through a USRP X310, while USRP B210 were used for the \glspl{ue}.
As this testbed does not natively support the \gls{lxc} virtualization technology, the \openrangym container images were transferred from Colosseum to the compute nodes through the \texttt{scp} utility, instantiated on Ubuntu Linux images loaded on the bare-metal servers of the testbed.

\textbf{COSMOS.}
%
%
In this case, the near-RT \gls{ric} was instantiated on a Supermicro 1028U-TRT+ server.
%
Base station and \gls{ue}, instead, were virtualized on Asus servers
driving USRP B210 \glspl{sdr}.
Similarly to what done for \gls{powder}, as the \gls{lxc} virtualization technology is not directly supported by this testbed, the container images were transferred from Colosseum through the \texttt{scp} utility, and instantiated on Ubuntu Linux images loaded on the bare-metal nodes available on the testbed.

\textbf{Arena.}
%
%
All applications were run on Dell PowerEdge R340 servers.
%
In this case, the \openrangym \gls{lxc} containers are instantiated directly on the bare-metal nodes of the testbed, which leverage USRP X310 \glspl{sdr} as radio front-ends.
On this testbed, the \glspl{ue} are implemented through commercial smartphones.

\textbf{Colosseum.}
To mimic the same deployment scenario used in the other testbeds, in Colosseum we considered cellular nodes deployed in a static \gls{rf} scenario without user mobility.
In this case the \gls{lxc} containers of \gls{ric}, base station, and \glspl{ue} directly run on Colosseum bare-metal Dell PowerEdge R730 servers.
%
All the cellular nodes leverage USRP X310 \glspl{sdr} as radio-front ends.

\subsection{Results}

To showcase the flexibilty of \openrangym in dynamically reconfiguring the spectrum allocated to the network slices across different testbeds, Figures~\ref{fig:slice-stairs} and~\ref{fig:slice-v} show the overall throughput of each network slice varying the resources allocated to them, in terms of \glspl{rbg}.
%
95\% confidence intervals are also represented by the shaded areas in the figures.
For both figures, the percentage of \glspl{rbg} allocated to each slice of the base station---which uses a $10$\:MHz configuration---is dynamically changed through the \scope \glspl{api} according to the following configuration (also summarized in Table~\ref{tab:slice-configuration}).
\begin{table}[ht]
\setlength\belowcaptionskip{5pt}
    \centering
    \footnotesize
    \setlength{\tabcolsep}{2pt}
    \caption{Slicing configuration, expressed as percentage of \glspl{rbg}, used in Figures~\ref{fig:slice-stairs} and~\ref{fig:slice-v}.}
    \label{tab:slice-configuration}
    \begin{tabularx}{\columnwidth}{
        >{\raggedright\arraybackslash\hsize=0.8\hsize}X
        >{\raggedright\arraybackslash\hsize=0.6\hsize}X
        >{\centering\arraybackslash\hsize=1.2\hsize}X
        >{\centering\arraybackslash\hsize=1.2\hsize}X
        >{\centering\arraybackslash\hsize=1.2\hsize}X }
        \toprule
        Figure & Slice & First Minute & Second Minute & Third Minute \\
        \midrule
        \multirow{2}*{Figure~\ref{fig:slice-stairs}} & Slice~A   & 75\%~\glspl{rbg}  & 50\%~\glspl{rbg} & 25\%~\glspl{rbg} \\
         & Slice~B   & 25\%~\glspl{rbg}   & 50\%~\glspl{rbg} & 75\%~\glspl{rbg} \\
        \midrule
        \multirow{2}*{Figure~\ref{fig:slice-v}} & Slice~A   & 75\%~\glspl{rbg}  & 25\%~\glspl{rbg}   & 75\%~\glspl{rbg} \\
         & Slice~B   & 25\&~\glspl{rbg}   & 75\%~\glspl{rbg}  & 25\%~\glspl{rbg} \\
        \bottomrule
    \end{tabularx}
\end{table}
%
%
In Figure~\ref{fig:slice-stairs}, the two network slices, i.e., slice~A and~B in the figure, are allocated the following \glspl{rbg} percentage: (i)~75\% to slice~A and 25\% to slice~B in the first minute; (ii)~50\% to each slice in the second minute, and (iii)~25\% to slice~A and 75\% to slice~B in the third minute.
In Figure~\ref{fig:slice-v}, instead they are allocated the following \glspl{rbg} percentage: (i)~75\% to slice~A and 25\% to slice~B in the first minute; (ii)~25\% to slice~A and 75\% to slice~B in the second minute, and (iii)~75\% to slice~A and 25\% to slice~B in the third minute.
In both these figures, the throughput varies proportionally to the specific allocation of slice resources, in which slices with more \glspl{rbg} achieve higher throughput values.
These values then change during the experiment as \glspl{rbg} are dynamically reallocated to the slices.
We notice that even if the throughput differs across the various testbeds because of the different capabilities and environments they offer---with Arena achieving the highest performance due to the use of commercial smartphones as the \glspl{ue}---the overall trends are consistent across the different setups.

\begin{figure}[ht]
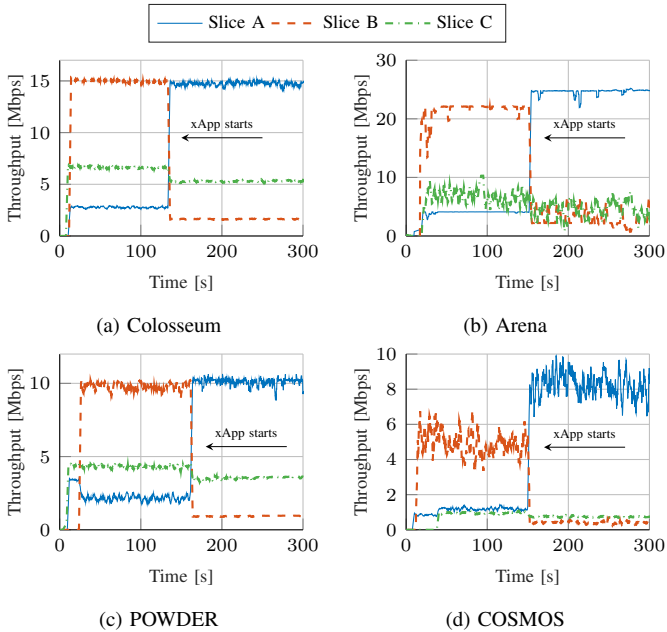

  \centering
  \ifexttikz
      \tikzsetnextfilename{ric-colosseum}
  \fi
  \begin{subfigure}[t]{0.48\columnwidth}
    \setlength\fwidth{.8\columnwidth}
    \setlength\fheight{.55\columnwidth}
    \input{figures/ric-colosseum-109.tex}
    \caption{Colosseum}
    \label{fig:ric-colosseum}
  \end{subfigure}\hfill
  \ifexttikz
      \tikzsetnextfilename{ric-arena}
  \fi
  \begin{subfigure}[t]{0.48\columnwidth}
    \setlength\fwidth{.8\columnwidth}
    \setlength\fheight{.55\columnwidth}
    \input{figures/ric-arena-106.tex}
    \caption{Arena}
    \label{fig:ric-arena}
  \end{subfigure}\hfill
  \ifexttikz
      \tikzsetnextfilename{ric-powder}
  \fi
  \begin{subfigure}[t]{0.48\columnwidth}
    \setlength\fwidth{.8\columnwidth}
    \setlength\fheight{.55\columnwidth}
    \input{figures/ric-powder-116.tex}
    \caption{POWDER}
    \label{fig:ric-powder}
  \end{subfigure}\hfill
  \ifexttikz
      \tikzsetnextfilename{ric-cosmos}
  \fi
  \begin{subfigure}[t]{0.48\columnwidth}
    \setlength\fwidth{.8\columnwidth}
    \setlength\fheight{.55\columnwidth}
    \input{figures/ric-cosmos-59.tex}
    \caption{COSMOS}
    \label{fig:ric-cosmos} 
  \end{subfigure}
  \caption{Slice throughput when the \scope \gls{ran} is controlled by \coloran near-RT \gls{ric}. At around second $150$, xApp to prioritize the amount of resources (i.e., \glspl{rbg}) allocated to slice~A is instantiated on the near-RT \gls{ric}.}
  \label{fig:ric-testbeds}
\end{figure}

We now showcase an instance in which the \coloran near-RT \gls{ric} is leveraged to control a softwarized \gls{ran} implemented through \scope.
\gls{lxc} containers for both applications are deployed on the testbeds mentioned above, whose specifications are summarized in Table~\ref{tab:compute-node-setup}.
Figure~\ref{fig:ric-testbeds} shows the evolution in time of the throughput of the three network slices (namely, slice~A, B, and~C) implemented by the \scope base station.
Initially, we consider the baseline configuration where no control is performed by the \gls{ric}.
Then, at around second $150$, an xApp that prioritizes one of the network slices (slice~A in the figure) is instantiated on the near-RT \gls{ric}.
Following the O-RAN \gls{ran} slicing \gls{sm}, this is done by reconfiguring at run time the number of \glspl{rbg} that are assigned exclusively to each slice, and can be used to schedule transmissions for users belonging to that slice~\cite{oran-wg3-e2-sm}. This, for example, allows slice~A to transmit across many \glspl{rbg}, as shown in Figure~\ref{fig:ric-testbeds}.
As a result, the xApp dynamically reallocates the amount of \glspl{rbg} of each slice, which reflects on the performance of the slices of the \gls{ran}.
Similar to the previous case, slices with a larger amount of \glspl{rbg} allocated to them achieve higher throughput values.
Overall, even in this case results are consistent across the different testbeds.

Now we show some timing statistics on the average amount of time taken to transfer the \scope and \coloran \gls{lxc} images from Colosseum to the Arena, \gls{cosmos}, and \gls{powder} platforms.
All the image transfers were performed through the \texttt{scp} utility, while the \gls{lxc} containers were created following the procedures detailed in Section~\ref{sec:traveling}.
In both cases, these timing statistics were derived using the hardware of Table~\ref{tab:compute-node-setup}.
We used the compute nodes listed in the ``base station (BS)/UE'' section of the table for the \scope \gls{lxc} image/container (in the case of the \gls{powder} platform, in which different compute nodes were are listed for base station and \gls{ue}, the base station node was used), and the compute nodes in ``near-RT RIC'' section of the table for \coloran.
In the tables that will be described next, we consider the following \gls{lxc} images:
\begin{itemize}
    \item \textit{\scope w/ E2}: this is the \scope \gls{lxc} image with the O-RAN E2 termination to interface with the near-RT \gls{ric}.
    \item \textit{\coloran near-RT \gls{ric}, prebuilt}: this is the \coloran \gls{lxc} image in which the Docker containers of the \gls{ric}, and sample xApp (described in Section~\ref{sec:coloran}) have been built a priori.
    \item \textit{\coloran near-RT \gls{ric}, to build}: this is the \coloran \gls{lxc} image with the scripts to build the Docker containers of the \gls{ric} and sample xApp from scratch. 
\end{itemize}
%
%
Table~\ref{tab:scp-transfer-time} shows the average time required to transfer the \gls{lxc} images from Colosseum to the other platforms.
\begin{table}[ht]
\setlength\belowcaptionskip{5pt}
    \centering
    \footnotesize
    \setlength{\tabcolsep}{2pt}
    \caption{Average time to transfer the LXC images from Colosseum to specific testbeds. The size of each image is listed in brackets.}
    \label{tab:scp-transfer-time}
    \begin{tabularx}{\columnwidth}{
        >{\raggedright\arraybackslash\hsize=0.6\hsize}X
        >{\raggedright\arraybackslash\hsize=0.8\hsize}X
        >{\raggedright\arraybackslash\hsize=1.3\hsize}X
        >{\raggedright\arraybackslash\hsize=1.3\hsize}X }
        \toprule
        Testbed & \scope w/ E2 ($1.7$\:GB) & \coloran near-RT RIC, prebuilt ($6.5$\:GB) & \coloran near-RT RIC, to build ($1.6$\:GB) \\
        \midrule
        Arena   & $1$\:m $27.413$\:s & $5$\:m $41.487$\:s & $1$\:m $25.002$\:s \\
        COSMOS  & $1$\:m $28.631$\:s & $5$\:m $39.704$\:s & $1$\:m $27.352$\:s \\
        POWDER  & $1$\:m $30.787$\:s & $5$\:m $43.704$\:s & $1$\:m $28.546$\:s \\
        \bottomrule
    \end{tabularx}
\end{table}
Times span from as low as $\sim 1.5$\:minutes to as high as almost $6$\:minutes, depending on the size of each image---also listed in the table---and capabilities of the testbeds.
However, transfer times are consistent across the different testbeds.

Finally, Table~\ref{tab:container-start-time} shows the times taken to instantiate \gls{lxc} containers from the images transferred from Colosseum.
\begin{table}[ht]
\setlength\belowcaptionskip{5pt}
    \centering
    \footnotesize
    \setlength{\tabcolsep}{2pt}
    \caption{Average time to start as a container the LXC image exported from Colosseum on specific testbeds. The size of each image is listed in brackets.}
    \label{tab:container-start-time}
    \begin{tabularx}{\columnwidth}{
        >{\raggedright\arraybackslash\hsize=0.6\hsize}X
        >{\raggedright\arraybackslash\hsize=0.8\hsize}X
        >{\raggedright\arraybackslash\hsize=1.3\hsize}X
        >{\raggedright\arraybackslash\hsize=1.3\hsize}X }
        \toprule
        Testbed & \scope w/ E2 ($1.7$\:GB) & \coloran near-RT RIC, prebuilt ($6.5$\:GB) & \coloran near-RT RIC, to build ($1.6$\:GB) \\
        \midrule
        Arena   & $0.887$\:s   & $1$\:m $11.483$\:s   & $46$\:m $18.110$\:s \\
        COSMOS  & $25.463$\:s  & $2$\:m $34.905$\:s   & $26$\:m $4.410$\:s \\
        POWDER  & $30.139$\:s  & $2$\:m $55.654$\:s   & $21$\:m $11.220$\:s \\
        \bottomrule
    \end{tabularx}
\end{table}
In this case, we notice some difference among the times achieved on the different testbeds.
For instance, Arena is significantly faster than \gls{cosmos} and \gls{powder} in instantiating the \scope container---completing the instantiation in less than $1$\:s---and the prebuilt \coloran container (instantiation in approximately $1$\:minute).
This is mainly due to the fact that Arena allows users to instantiate applications on the bare-metal nodes directly.
This removes the latency of the extra virtualization layer of the other two testbeds, in which the \gls{lxc} containers are nested inside the virtualized architecture the users are given access to.
%
%
When it comes to building the Docker containers of the \coloran near-RT \gls{ric} (see Section~\ref{sec:coloran}) from scratch, instead, \gls{powder} and \gls{cosmos} are significantly faster than Arena, taking approximately half the time to complete the same operations.
This is mainly due to the superior compute capabilities of the nodes of these two testbeds (24~core CPU server on \gls{powder}, and 16~core server on \gls{cosmos} vs.\ 6~core CPU server on Arena).
Nonetheless, this building operation needs to be completed only once, as the compiled \coloran \gls{lxc} image can be saved to be used in subsequent experiments, with instantiation times sensibly lower (slightly above $1$\:minute for Arena, and below $3$\:minutes for \gls{powder} and \gls{cosmos}). 

\section{Conclusions}
\label{sec:conclusions}

We presented \openrangym, the first publicly-available research platform for data-driven O-RAN experimentation at scale on heterogeneous wireless testbeds.
Building on, and extending, frameworks for data collection and \gls{ran} control, \openrangym enables the end-to-end design and testing of data-driven xApps instantiated on the O-RAN infrastructure.
We described the core components of \openrangym---including frameworks and experimental platforms---and detailed procedures and configuration options for experimenting at scale on a softwarized \gls{ran} instantiated on Colosseum.
Then, we gave an overview of the xApp design and testing workflow enabled by \openrangym, also showcasing an example of two xApps
used to control a large-scale O-RAN managed softwarized \gls{ran} deployed on Colosseum.
Finally, we demonstrated how \openrangym solutions and experiments can be transitioned from Colosseum to heterogeneous real-world platforms, such as the Arena testbed, and the POWDER and COSMOS platforms of the PAWR program.
\openrangym is publicly-available to the research community, and opened up for community contributions and additions.

\footnotesize  
\bibliographystyle{IEEEtran}
\bibliography{biblio.bib}

\begin{thebibliography}{10}
\providecommand{\url}[1]{#1}
\csname url@samestyle\endcsname
\providecommand{\newblock}{\relax}
\providecommand{\bibinfo}[2]{#2}
\providecommand{\BIBentrySTDinterwordspacing}{\spaceskip=0pt\relax}
\providecommand{\BIBentryALTinterwordstretchfactor}{4}
\providecommand{\BIBentryALTinterwordspacing}{\spaceskip=\fontdimen2\font plus
\BIBentryALTinterwordstretchfactor\fontdimen3\font minus
  \fontdimen4\font\relax}
\providecommand{\BIBforeignlanguage}[2]{{%
\expandafter\ifx\csname l@#1\endcsname\relax
\typeout{** WARNING: IEEEtran.bst: No hyphenation pattern has been}%
\typeout{** loaded for the language `#1'. Using the pattern for}%
\typeout{** the default language instead.}%
\else
\language=\csname l@#1\endcsname
\fi
#2}}
\providecommand{\BIBdecl}{\relax}
\BIBdecl

\bibitem{bonati2022openrangym}
L.~Bonati, M.~Polese, S.~D'Oro, S.~Basagni, and T.~Melodia, ``{OpenRAN Gym: An
  Open Toolbox for Data Collection and Experimentation with AI in O-RAN},'' in
  \emph{Proceedings of IEEE WCNC Workshop on Open RAN Architecture for 5G
  Evolution and 6G}, Austin, TX, USA, April 2022.

\bibitem{bonati2020open}
L.~Bonati, M.~Polese, S.~{D’Oro}, S.~Basagni, and T.~Melodia, ``Open,
  programmable, and virtualized {5G} networks: State-of-the-art and the road
  ahead,'' \emph{Computer Networks}, vol. 182, pp. 1--28, December 2020.

\bibitem{oran-wg1-arch-spec}
{O-RAN Working Group 1}, ``{{O-RAN} Architecture Description 5.00},''
  O-RAN.WG1.O-RAN-Architecture-Description-v05.00 Technical Specification, July
  2021.

\bibitem{polese2022understanding}
M.~Polese, L.~Bonati, S.~D'Oro, S.~Basagni, and T.~Melodia, ``{Understanding
  O-RAN: Architecture, Interfaces, Algorithms, Security, and Research
  Challenges},'' \emph{arXiv:2202.01032 [cs.NI]}, February 2022.

\bibitem{oran-wg3-ricarch}
{O-RAN Working Group 3}, ``{O-RAN Near-RT RAN Intelligent Controller Near-RT
  RIC Architecture 2.00},'' O-RAN.WG3.RICARCH-v02.00, March 2021.

\bibitem{oran-wg2-non-rt-ric-architecture}
{O-RAN Working Group 2}, ``{{O-RAN} Non-RT RIC Architecture 1.0},''
  O-RAN.WG2.Non-RT-RIC-ARCH-TS-v01.00 Technical Specification, July 2021.

\bibitem{oran-wg2-ml}
------, ``{O-RAN AI/ML} workflow description and requirements 1.03,''
  O-RAN.WG2.AIML-v01.03 Technical Specification, July 2021.

\bibitem{polese2021coloran}
M.~Polese, L.~Bonati, S.~D'Oro, S.~Basagni, and T.~Melodia, ``{ColO-RAN:
  Developing Machine Learning-based xApps for Open RAN Closed-loop Control on
  Programmable Experimental Platforms},'' \emph{IEEE Transactions on Mobile
  Computing}, pp. 1--14, July 2022.

\bibitem{doro2022dapps}
S.~D'Oro, M.~Polese, L.~Bonati, H.~Cheng, and T.~Melodia, ``{dApps: Distributed
  Applications for Real-time Inference and Control in O-RAN},'' \emph{IEEE
  Communications Magazine}, pp. 1--7, 2022, in print; preprint available at
  \url{https://arxiv.org/pdf/2203.02370.pdf}.

\bibitem{bonati2021colosseum}
L.~Bonati, P.~Johari, M.~Polese, S.~D'Oro, S.~Mohanti, M.~Tehrani-Moayyed,
  D.~Villa, S.~Shrivastava, C.~Tassie, K.~Yoder, A.~Bagga, P.~Patel, V.~Petkov,
  M.~Seltser, F.~Restuccia, A.~Gosain, K.~R. Chowdhury, S.~Basagni, and
  T.~Melodia, ``{Colosseum: Large-Scale Wireless Experimentation Through
  Hardware-in-the-Loop Network Emulation},'' in \emph{Proceedings of IEEE
  DySPAN}, December 2021.

\bibitem{bonati2021scope}
L.~Bonati, S.~D'Oro, S.~Basagni, and T.~Melodia, ``{SCOPE}: An open and
  softwarized prototyping platform for {NextG} systems,'' in \emph{Proceedings
  of ACM MobiSys}, June 2021.

\bibitem{bertizzolo2020arena}
L.~Bertizzolo, L.~Bonati, E.~Demirors, A.~Al-Shawabka, S.~{D'Oro},
  F.~Restuccia, and T.~Melodia, ``{{Arena}: A 64-antenna {SDR}-based Ceiling
  Grid Testing Platform for Sub-6 {GHz} {5G}-and-Beyond Radio Spectrum
  Research},'' \emph{Computer Networks}, vol. 181, pp. 1--17, November 2020.

\bibitem{pawr}
{\acrfull{pawr}}. \url{https://www.advancedwireless.org}. Accessed December
  2021.

\bibitem{breen2021powder}
J.~Breen, A.~Buffmire, J.~Duerig, K.~Dutt, E.~Eide, A.~Ghosh, M.~Hibler,
  D.~Johnson, S.~K. Kasera, E.~Lewis \emph{et~al.}, ``{POWDER: Platform for
  Open Wireless Data-driven Experimental Research},'' \emph{Computer Networks},
  vol. 197, pp. 1--18, October 2021.

\bibitem{raychaudhuri2020cosmos}
D.~Raychaudhuri, I.~Seskar, G.~Zussman, T.~Korakis, D.~Kilper, T.~Chen,
  J.~Kolodziejski, M.~Sherman, Z.~Kostic, X.~Gu, H.~Krishnaswamy,
  S.~Maheshwari, P.~Skrimponis, and C.~Gutterman, ``{Challenge: {COSMOS}: A
  City-Scale Programmable Testbed for Experimentation with Advanced
  Wireless},'' in \emph{Proceedings of ACM MobiCom}, London, United Kingdom,
  Sept. 2020.

\bibitem{s21248173}
M.~Dryjanski, L.~Kulacz, and A.~Kliks, ``{Toward Modular and Flexible Open RAN
  Implementations in 6G Networks: Traffic Steering Use Case and O-RAN xApps},''
  \emph{Sensors}, vol.~21, no.~24, pp. 1--14, December 2021.

\bibitem{johnson2021open}
D.~Johnson, D.~Maas, and J.~Van Der~Merwe, ``{Open Source RAN Slicing on
  POWDER: A Top-to-Bottom O-RAN Use Case},'' in \emph{Proceedings of ACM
  MobiSys}, June 2021.

\bibitem{lee2020hosting}
H.~Lee, J.~Cha, D.~Kwon, M.~Jeong, and I.~Park, ``{Hosting AI/ML Workflows on
  O-RAN RIC Platform},'' in \emph{Proceedings of IEEE GLOBECOM Workshops},
  December 2020.

\bibitem{abdalla2021generation}
A.~S. Abdalla, P.~S. Upadhyaya, V.~K. Shah, and V.~Marojevic, ``{Toward Next
  Generation Open Radio Access Network--What O-RAN Can and Cannot Do!}''
  \emph{arXiv preprint arXiv:2111.13754 [cs.NI]}, November 2021.

\bibitem{oran2019plugfest}
{O-RAN Alliance Conducts First Global Plugfest to Foster Adoption of Open and
  Interoperable 5G Radio Access Networks}. (2019, December)
  \url{https://tinyurl.com/f48auynf}.

\bibitem{doro2022orchestran}
S.~D'Oro, L.~Bonati, M.~Polese, and T.~Melodia, ``{OrchestRAN: Network
  Automation through Orchestrated Intelligence in the Open RAN},'' in
  \emph{Proceedings of IEEE INFOCOM}, May 2022, arXiv:2201.05632 [cs.NI].

\bibitem{gomez2016srslte}
I.~Gomez-Miguelez, A.~Garcia-Saavedra, P.~Sutton, P.~Serrano, C.~Cano, and
  D.~Leith, ``{srsLTE}: An open-source platform for {LTE} evolution and
  experimentation,'' in \emph{Proceedings of ACM WiNTECH}, October 2016.

\bibitem{kaltenberger2020openairinterface}
F.~Kaltenberger, A.~P.~Silva, A.~Gosain, L.~Wang, and T.-T. Nguyen,
  ``{OpenAirInterface}: Democratizing innovation in the {5G} era,''
  \emph{Computer Networks}, no. 107284, May 2020.

\bibitem{kohli2021openaccess}
M.~Kohli, T.~Chen, M.~B. Dastjerdi, J.~Welles, I.~Seskar, H.~Krishnaswamy, and
  G.~Zussman, ``{Open-Access Full-Duplex Wireless in the ORBIT and COSMOS
  Testbeds},'' \emph{Computer Networks}, 2021.

\bibitem{bonati2021intelligence}
L.~Bonati, S.~D'Oro, M.~Polese, S.~Basagni, and T.~Melodia, ``{Intelligence and
  Learning in O-RAN for Data-driven NextG Cellular Networks},'' \emph{IEEE
  Communications Magazine}, vol.~59, no.~10, pp. 21--27, October 2021.

\bibitem{osc-du-l2}
{O-RAN Software Community}. {O-DU L2 Repository}.
  \url{https://github.com/o-ran-sc/o-du-l2}. Accessed December 2021.

\bibitem{johnson-powder-ric-profile}
{David Johnson}. {POWDER RIC Profile Repository}.
  \url{https://gitlab.flux.utah.edu/johnsond/ric-profile}. Accessed December
  2021.

\bibitem{bonati2022intelligent}
L.~Bonati, M.~Polese, S.~D'Oro, S.~Basagni, and T.~Melodia, ``{Intelligent
  Closed-loop RAN Control with xApps in OpenRAN Gym},'' in \emph{Proceedings of
  European Wireless 2022}, Dresden, Germany, September 2022.

\bibitem{oran-wg3-e2-sm}
{O-RAN Working Group 3}, ``{O-RAN} near-real-time {RAN} intelligent controller
  {E2} service model 2.00,'' ORAN-WG3.E2SM-v02.00 Technical Specification, July
  2021.

\end{thebibliography}

\begin{IEEEbiography}
[{\includegraphics[width=1in,height=1.25in,keepaspectratio]{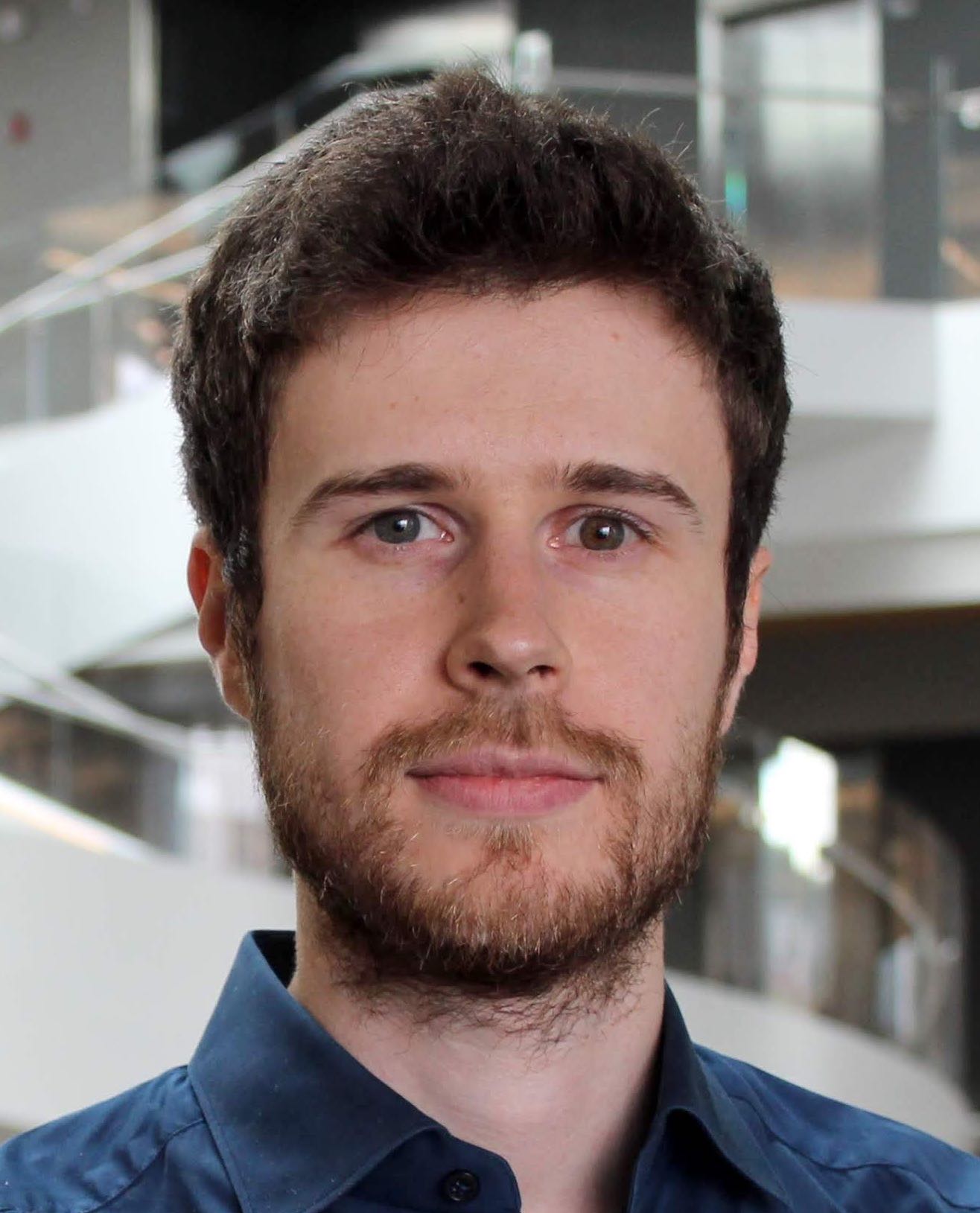}}]{Leonardo Bonati}
is an Associate Research Scientist at the Institute for the Wireless Internet of Things, Northeastern University, Boston, MA. He received the Ph.D. degree in Computer Engineering from Northeastern University in 2022. His main research focuses on softwarized approaches for the Open Radio Access Network (RAN) of the next generation of cellular networks, on O-RAN-managed networks, and on network automation and orchestration. He served multiple times on the technical program committee of the ACM Workshop on Wireless Network Testbeds, Experimental evaluation \& Characterization, and as guest editor of the special issue of Elsevier's Computer Networks journal on Advances in Experimental Wireless Platforms and Systems.
\end{IEEEbiography}


\begin{IEEEbiography}
[{\includegraphics[width=1in,height=1.25in,keepaspectratio]{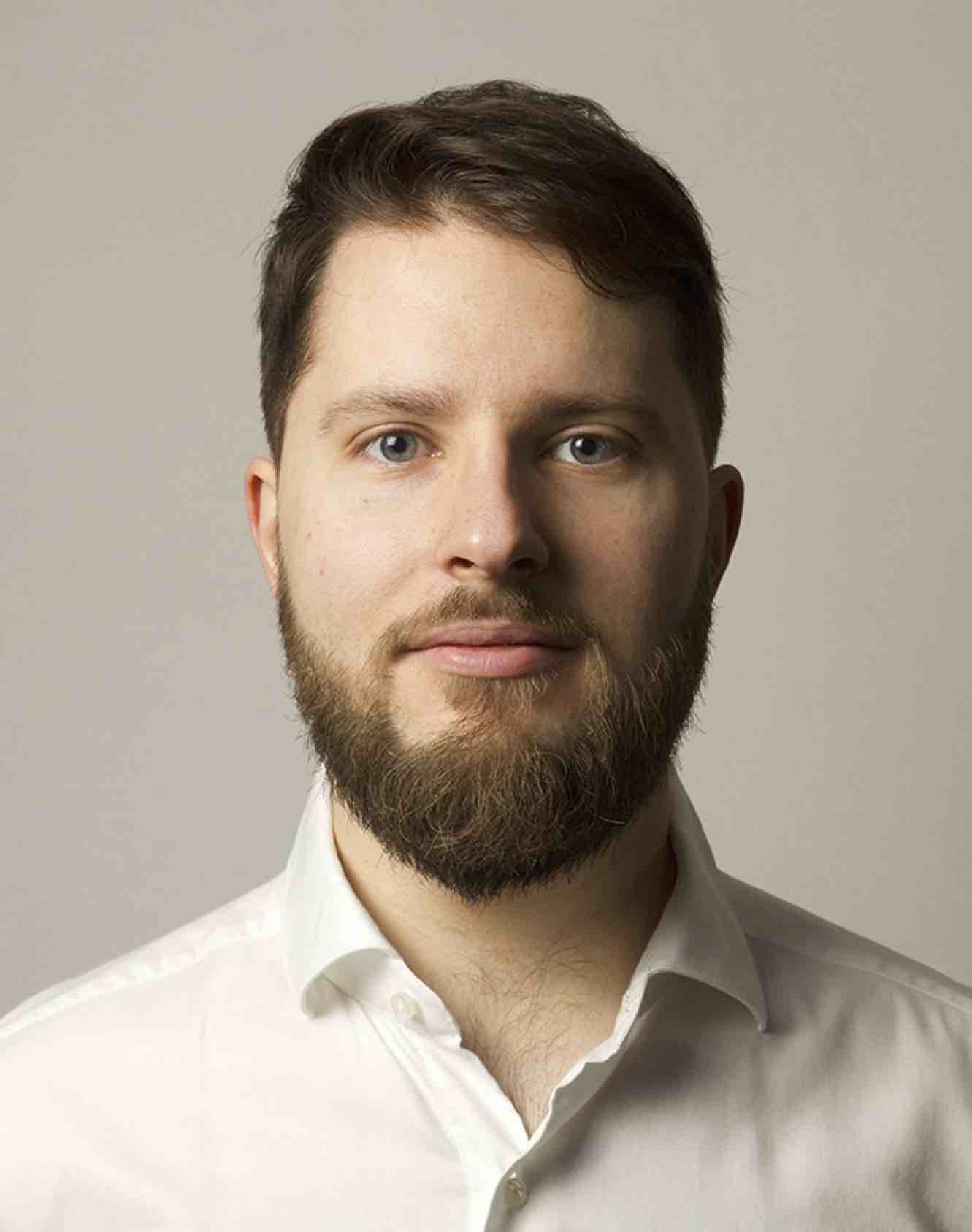}}]{Michele Polese} is a Principal Research Scientist at the Institute for the Wireless Internet of Things, Northeastern University, Boston, since March 2020. He received his Ph.D. at the Department of Information Engineering of the University of Padova in 2020. He also was an adjunct professor and postdoctoral researcher in 2019/2020 at the University of Padova, and a part-time lecturer in Fall 2020 and 2021 at Northeastern University. During his Ph.D., he visited New York University (NYU), AT\&T Labs in Bedminster, NJ, and Northeastern University.
His research interests are in the analysis and development of protocols and architectures for future generations of cellular networks (5G and beyond), in particular for millimeter-wave and terahertz networks, spectrum sharing and passive/active user coexistence, open RAN development, and the performance evaluation of end-to-end, complex networks. He has contributed to O-RAN technical specifications and submitted responses to multiple FCC and NTIA notice of inquiry and requests for comments, and is a member of the Committee on Radio Frequency Allocations of the American Meteorological Society (2022-2024). He collaborates and has collaborated with several academic and industrial research partners, including AT\&T, Mavenir, NVIDIA, InterDigital, NYU, University of Aalborg, King's College, and NIST. He was awarded with several best paper awards, is serving as TPC co-chair for WNS3 2021-2022, as an Associate Technical Editor for the IEEE Communications Magazine, and has organized the Open 5G Forum in Fall 2021. He is a Member of the IEEE.
\end{IEEEbiography}


\begin{IEEEbiography}
[{\includegraphics[width=1in,height=1.25in,keepaspectratio]{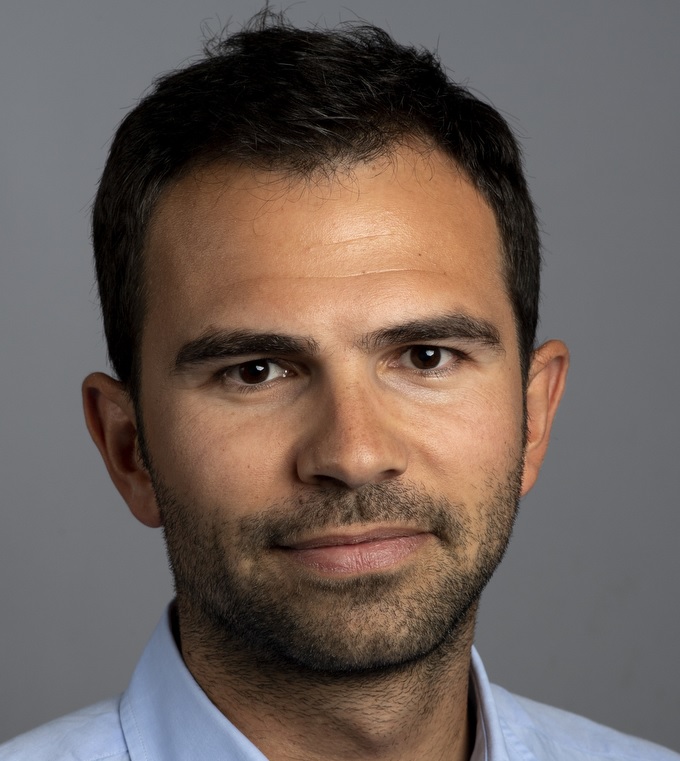}}]{Salvatore D'Oro}
is a Research Assistant Professor at Northeastern University. He received his Ph.D. degree from the University of Catania in 2015. Salvatore is an area editor of Elsevier Computer Communications journal and serves on the Technical Program Committee (TPC) of multiple conferences and workshops such as IEEE INFOCOM, IEEE CCNC, IEEE ICC and IFIP Networking. Dr. D'Oro's research interests include optimization, artificial intelligence, security, network slicing and their applications to 5G networks and beyond. He is a Member of the IEEE.
\end{IEEEbiography}


\begin{IEEEbiography}
[{\includegraphics[width=1in,height=1.25in,keepaspectratio]{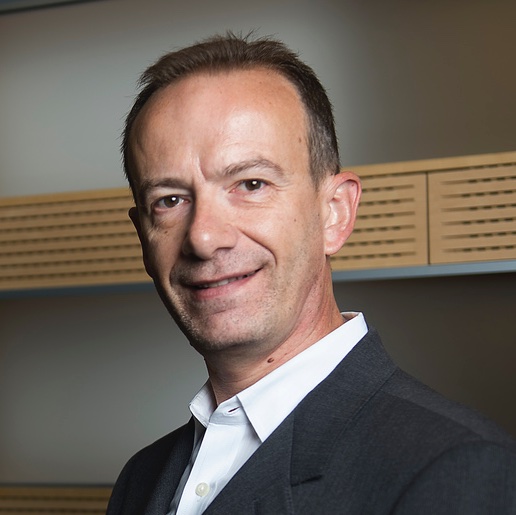}}]{Stefano Basagni}
is with the Institute for the Wireless Internet of Things and a professor at the ECE Department at Northeastern University, in Boston, MA. He holds a Ph.D.\ in electrical engineering from the University of Texas at Dallas (2001) and a Ph.D.\ in computer science from the University of Milano, Italy (1998). Dr. Basagni's current interests concern research and implementation aspects of mobile networks and wireless communications systems, wireless sensor networking for IoT (underwater, aerial and terrestrial), and definition and performance evaluation of network protocols.
Dr. Basagni has published over twelve dozen of highly cited, refereed technical papers and book chapters. His h-index is currently 49 (November 2022). He is also co-editor of three books. Dr. Basagni served as a guest editor of multiple international ACM/IEEE, Wiley and Elsevier journals. He has been the TPC co-chair of international conferences. He is a distinguished scientist of the ACM, a senior member of the IEEE, and a member of CUR (Council for Undergraduate Education).
\end{IEEEbiography}

\begin{IEEEbiography}
[{\includegraphics[width=1in,height=1.25in,keepaspectratio]{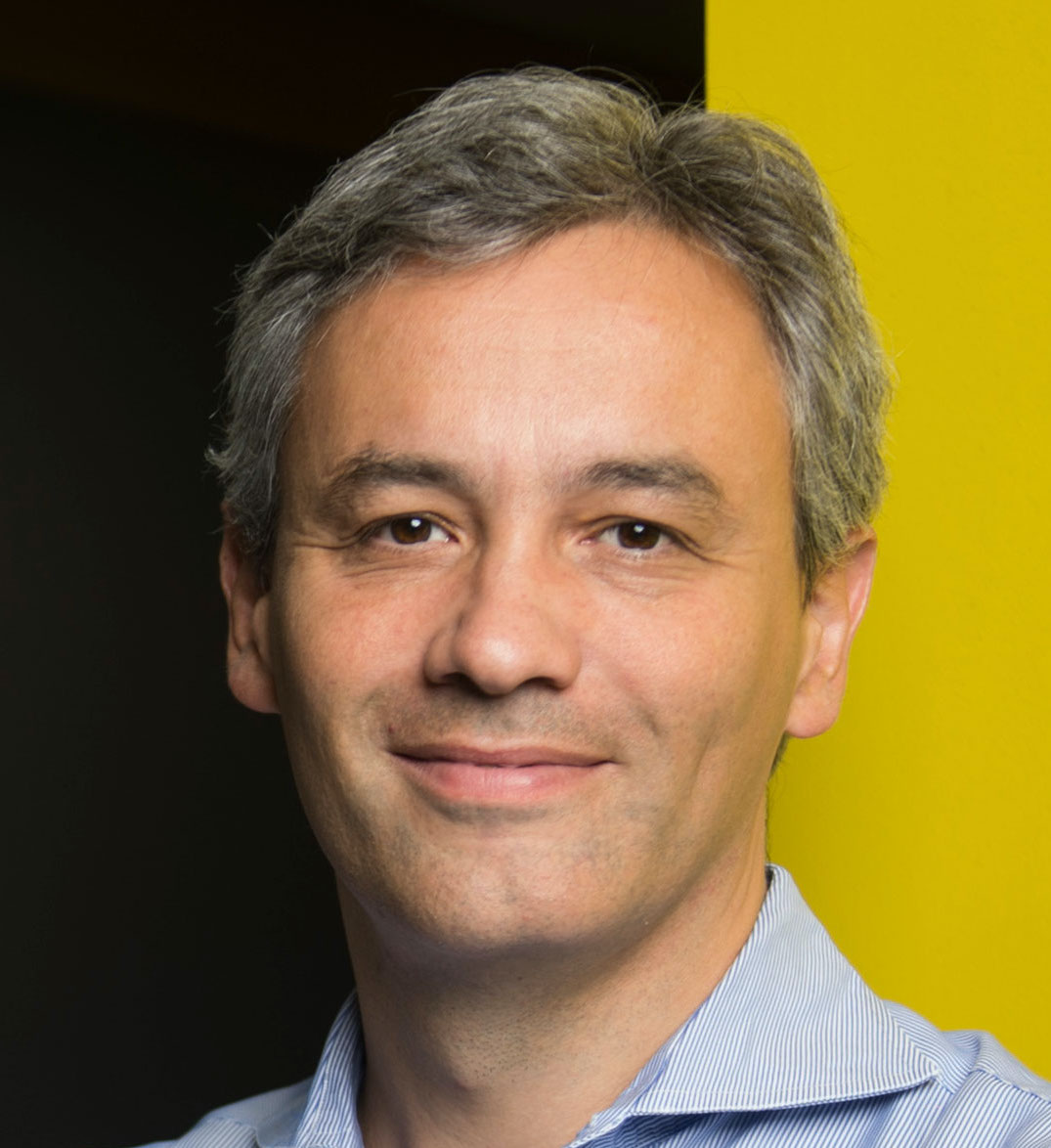}}]{Tommaso Melodia}
is the William Lincoln Smith Chair Professor with the Department of Electrical and Computer Engineering at Northeastern University in Boston. He is also the Founding Director of the Institute for the Wireless Internet of Things and the Director of Research for the PAWR Project Office. He received his Ph.D. in Electrical and Computer Engineering from the Georgia Institute of Technology in 2007. He is a recipient of the National Science Foundation CAREER award. Prof. Melodia has served as Associate Editor of IEEE Transactions on Wireless Communications, IEEE Transactions on Mobile Computing, Elsevier Computer Networks, among others. He has served as Technical Program Committee Chair for IEEE Infocom 2018, General Chair for IEEE SECON 2019, ACM Nanocom 2019, and ACM WUWnet 2014. Prof. Melodia is the Director of Research for the Platforms for Advanced Wireless Research (PAWR) Project Office, a \$100M public-private partnership to establish 4 city-scale platforms for wireless research to advance the US wireless ecosystem in years to come. Prof. Melodia's research on modeling, optimization, and experimental evaluation of Internet-of-Things and wireless networked systems has been funded by the National Science Foundation, the Air Force Research Laboratory the Office of Naval Research, DARPA, and the Army Research Laboratory. Prof. Melodia is a Fellow of the IEEE and a Senior Member of the ACM.
\end{IEEEbiography}

\vfill

\end{document}